\documentclass{aa}

\usepackage{natbib}
\bibpunct{(}{)}{;}{a}{}{,} 
\usepackage{graphics}   
\usepackage{graphicx}   
\usepackage{epstopdf}
\usepackage{longtable}   
\usepackage{url}      
\usepackage{bm}        
\usepackage[varg]{txfonts}
\usepackage{pdflscape}
\usepackage{supertabular}
\usepackage{hyperref}   
\usepackage{wasysym}
\usepackage[svgnames]{xcolor}

\hypersetup{
    unicode=false,          
    colorlinks=true,       
    linkcolor=blue,          
    citecolor=blue,        
    filecolor=blue,      
    urlcolor=blue           
}

\usepackage{color}

\begin{document}

\title{Star-planet interactions} 
\subtitle{VI. Tides, stellar activity and  planet evaporation}

\author{Suvrat Rao\inst{1,2,3}, Camilla Pezzotti\inst{1}, Georges Meynet\inst{1}, Patrick Eggenberger\inst{1}, Ga\"el Buldgen\inst{1}, Christoph Mordasini\inst{4}, Vincent Bourrier\inst{1}, Sylvia Ekstr{\"o}m\inst{1}, Cyril Georgy\inst{1}
}

\authorrunning{Rao et al.}

\institute{Geneva Observatory, University of Geneva, Maillettes 51, CH-1290 Sauverny, Switzerland
\and Indian Institute of Technology Kharagpur, 721302 West Bengal, India
\and Hamburg Observatory, University of Hamburg, Gojenbergsweg 112, 21029 Hamburg, Germany
\and 
Physikalisches Institut, University of Bern, Sidlerstrasse 5, CH-3012 Bern, Switzerland}

\date{Received /
Accepted}
\abstract
{Tidal interactions and planet evaporation processes impact the evolution of close-in star-planet systems.
}
{We study the impact of stellar rotation on these processes.
}
{We compute the time evolution of star-planet systems consisting of a planet with initial mass between 0.02 and 2.5 M$_{ Jup}$ (6 and 800 M$_{ Earth}$), in a quasi-circular orbit with an initial orbital distance between 0.01 and 0.10 au, around a solar-type star evolving from the Pre-Main-Sequence (PMS) until the end of the Main-Sequence (MS) phase. We account for the evolution of: the stellar structure, the stellar angular momentum due to tides and magnetic braking, the tidal interactions (equilibrium and dynamical tides in stellar convective zones), the mass-evaporation of the planet, and the secular evolution of the planetary orbit. We consider that at the beginning of the evolution, the proto-planetary disk has fully dissipated and planet formation is complete.
}
{Both a rapid initial stellar rotation, and a more efficient angular momentum transport inside the star, in general, contribute toward the enlargement of the domain which is devoid of planets after the PMS phase, in the plane of planet mass vs. orbital distance. Comparisons with the observed distribution of exoplanets orbiting solar mass stars, in the plane of planet mass vs. orbital distance (addressing the "Neptunian desert" feature), show an encouraging agreement with the present simulations, especially since no attempts have been made to fine-tune initial parameters of the models to fit the observations. We also obtain an upper limit for the orbital period of bare-core planets, that agrees with observations of the "radius valley" feature in the plane of planet radius vs. orbital period. 
}
{
The two effects, tides and planet evaporation, should be accounted for simultaneously and in a consistent way, with a detailed model for the evolution of the star. 
}


\maketitle

\titlerunning{Planet evaporation}
\authorrunning{Rao et al.}

\section{Introduction}

Numerous exoplanets, ranging from small rocky objects to planets more massive than Jupiter, have been discovered orbiting very close to their host (see the database http://exoplanet.eu). 
The vicinity between the planet and the star
affects and determines the various interactions between them, leading to significant consequences for the evolution of the star-planet system: 
A planet can orbit so close to its host star (planet grazing), that it can cause gas ejections from the star \citep{stephan2020}. Such ejections have been reported for the red giant star V Hydrae \citep{sahai2016,salas2019}. The star may interact with the magnetic field of the planet,
giving birth to hot spots on its surface \citep[see e.g.][]{strugarek2019} and also inducing some magnetic activity in the planet. 
The stellar rotation is affected by
transfers of angular momentum between the planetary orbit and the star through equilibrium and dynamical tides \citep{zahn77, Ogilvie2007, Ogil2013, Mathis2015}.
The irradiation received by the planet from the star impacts its
equilibrium temperature and thus the hydrostatic structure of its envelope \citep{Yelle2004, baraffe04, bear11, Debrecht2020}.
The energy deposited by high energy photons from the star may lead to inflation of the planet atmosphere  \citep[as e.g. WASP-121b][]{Delrez2016,Bourrier2020} and even to the partial or total loss of the planetary atmosphere \citep{Ehrenreich2015}. 
Interestingly, in the future, using the JWST (James Webb Space Telescope), observations of the planet evaporation process through transmission spectroscopy during transits might allow inferring the composition of the interior of small rocky planets \citep{Bodman2018}.
Some observed features, such as the existence of the radius valley and of the Neptunian desert in the plane of planet radius vs. orbital period \citep[see e.g.][]{vaneylen2018} are interpreted as being shaped, to a great extent, by the planet evaporation process \citep{Lopez2013,Lopez2013_2,Owen2013,Jin2014,Jin2018,Owen2018,Owen2019,Owen2019_2}.
The distance between the star and planet is modified by the tides and also by changes of the planetary and stellar masses.
This may cause a planet engulfment at some time during the evolution of the system \citep{siess99I, siess99II,villaver07, nordhaus10, kunimoto11, mustill12, nordhaus13,GIOII,stephan2020}.
A planet engulfment can be accompanied by transient phenomena such as strong stellar winds, emissions in visible, UV and X-ray radiation \citep{metzger2012, stephan2020} that may last for durations of 100-1000 years. Some authors \citep{morris1981,soker1998,kim2017}
have also suggested that planet engulfment can cause non-spherical, dipole-shaped planetary nebulae. Planet engulfment can also spin-up the star \citep{GIOII, qureshi2018, stephan2020}. 


Many works have studied the impact of tides on planet orbits 
\citep[see e.g.][]{livio84, soker84, sackmann93, rasio1996, siess99I, siess99II, villaver07, sato08, villaver09, carlberg09, nordhaus10, kunimoto11, bear11, mustill12, nordhaus13, villaver14, GIOI, GIOII, GIOIII, GIOIV, Bol2017, Gallet2018}, without considering for the most close-in planets, the effects of planet evaporation. Other studies have considered planet evaporation but do not account for the impact of tides induced by the planet on the star \citep[the only tidal effect accounted for is the Roche lobe overflow from the planet, see e.g.][]{Penz2008,Kuro2014,Ionov2018}.
Note that in most of the cases, there is a reasonable justification for neglecting tidal interactions. Indeed, as is well known \citep[see e.g. the review of][and references therein]{Owen2019}, evaporation mostly affects low mass planets whose evaporation timescale is in general much shorter than the timescale for their orbit to be changed by tides. For the most massive planets, however, this neglect is less justified. In a few studies, both
the processes of evaporation and tides are accounted for \citep[see e.g. ][]{Jackson2016, Cam2018, Owen2018}. However, in these works, only equilibrium tides or Darwin’s classical theory of tides are considered, which are not valid for fast rotating stars where dynamical tides may dominate. 
Most studies also consider only the average XUV (X-ray and Ultra-Violet radiation) luminosity coming from the star, not accounting for the possibility that this XUV luminosity can vary with the initial rotation of the star, which itself follows a natural evolution that can also be modified due to tidal interaction in the case of more massive planets.

In the present work, we make first steps towards a more coherent and consistent approach, in considering simultaneously the evolution of the planet orbit and mass, together with the evolution of the star. At the moment, to our knowledge, very few works, if any,
account in a consistent way for the effects of tides and evaporation, including the evolution of the star, whose rotational evolution is impacted by these processes. In this first step, however, we consider a simple model for the planet, likely too schematic, but sufficient in our view to explore possible feedbacks between tides, evaporation and stellar rotation. Moreover, comparisons with observations and with more sophisticated models for the planet indicate that despite the simple planet models we use here, the results obtained do not appear unrealistic.

Since this paper is the sixth in a series, we briefly explain its position
in terms of physics included, relative to our past works.
In the first papers \citep[][]{GIOIII, GIOI,GIOII}, we considered the case of planets sufficiently distant from their host star, to have a tidal interaction only during the red giant phase. Only the case of equilibrium tides was considered. We showed that
planet engulfment may cause an increase in the surface velocity of red giants well above what can be achieved even by considering the most extreme cases of initial rotation and angular momentum transport efficiency. We confirmed that planet engulfment could be responsible for increasing significantly the lithium abundance in the envelope of red giant stars \citep[][]{GIOIV}, and made predictions of the surface magnetic field of red giant stars after spin-up by planet engulfment 
\citep{Meynet2017}. In \citet{rao2018}, we studied instead the cases of  close-in planets to the host star during its Pre-Main-Sequence phase, accounting for dynamical tides along with equilibrium tides. We concluded that the region where the planet can be engulfed depends sensitively on the rotation of the star.

In the present work, we add a new piece in our approach. We account for the possible evaporation of the planet. This will mainly affect, of course, planets at a distance close to their host star, and concern mainly the Pre- and early Main-Sequence phase, when the star is still rotating sufficiently rapidly to emit high luminosity XUV radiation.
As indicated above, we study the interactions between the following processes: changes of the planetary orbit due to equilibrium and dynamical tides induced by the planet on the star; changes of the planetary mass due to the XUV irradiation of the planet by the star; and changes of the stellar rotation rate due to the stellar interior evolution, magnetic wind braking, tides and planet engulfment. These three aspects are tightly linked.
To illustrate this, let us consider a planet sufficiently close to its star so that its distance from the host star shrinks due to tidal interactions ({\it i.e.} implying that the orbital distance is inferior to the corotation radius). This will have at least two obvious effects: firstly, it will increase the irradiation flux received by the planet closer to the star, and second, since the star will receive some angular momentum, it will rotate faster, and therefore its activity will be different from the one it would have if evolving in isolation. This would also change the XUV irradiation and therefore, the evaporation rate of the planetary atmosphere. The mass loss of the planet has, in turn, an impact on the evolution of the orbit, thus impacting also the amplitudes of the tidal interactions between the star and the planet. 

In this paper we want to answer, within the limits of the hypotheses made in the present work (see Sect.~\ref{sec:2}), the following questions:
\begin{itemize}
\item How does the stellar rotation/activity react to the changes of the orbit, and how do these changes affect the irradiation of the planet and the photo-evaporation of its atmosphere.\\
\item In the plane of initial orbital distance vs. initial mass of the planet, which are the regions where tides dominate, where planet evaporation dominates and where both play a significant role in impacting the planetary orbit and mass?\\
\item How does the initial rotation of the star impact the fate of the planets? \\
\end{itemize}

The paper is organized as follows:
Sect.~\ref{sec:2} introduces the physics of the models.
We discuss the interplay between tides and the evaporation process in Sect.~\ref{sec:3}. 
Sect.~\ref{sec:4} summarises the different types of evolution of the star-planet systems considered in this study, depending on the initial conditions. Comparison of the predictions of our present simulations with observations is done in Sect.~\ref{sec:5}. Advantages and limitations of the present approach are discussed in Sect.~\ref{sec:6}. The main results are synthesized in Sect.~\ref{sec:7}.

\section{Physics of our computations}\label{sec:2}

\subsection{Stellar models}\label{sec:2.1}

We use a 1 M$_\odot$ star model that has been computed by the GENEC code \citep{egg08}. The computation of the model started from the Pre-Main-Sequence phase
when its radius was about 1.8 R$_\odot$. The time to reach the Zero Age Main-Sequence from this point is 40 My.
A full description of this 1 M$_\odot$ star model in a graphical form is given in Fig. B.1 of ~\citet{rao2018}.

We consider that the star has, at all times during the Pre-Main-Sequence (PMS) and Main-Sequence (MS) phases, a solid body rotation. This seems to be a reasonable hypothesis in view of the rotation profile deduced in the Sun from helioseismology \citep{Garcia2007,Eggenberger2019} and seismic analyses of solar-like MS stars using Kepler data \citep{Nielsen2015,Benomar2015,Benomar2018} and $\gamma$Dor stars \citep{Saio2021}. Hence, the angular momentum of the star, $L_{st} $, in terms of its moment of inertia, $I_{st} $, and stellar rotation, $\omega_{st} $, is given by,
$$
    L_{st} = I_{st}\omega_{st}.
$$

We decouple the evolution of the structure of the star, and therefore, of $I_{st}$ (by computing it with our stellar model independently for a given choice of the initial rotation), from the evolution of its rotation, $\omega_{st}$, which depends on the changes in the stellar interior structure, tidal interactions and magnetic interactions. We assume, therefore, that the feedback of rotation on the internal structure of the star is negligible. This is very well justified as long as the rotation is below about 70\% of the critical velocity at any point in the star. It happens that this condition is never violated in our case, or if so, only at the surface and
for a very short time. Thus, we can confidently say that indeed the structure evolution of the star can be taken the same for a very large range of rotations.

Overall, the equations describing the evolution of the angular momentum of the star, $L_{st} $, as a function of time are as follows:

\begin{equation}
 \dot{L}_{st} =  \dot{L}_{wb} + \dot{L}_{ti},
\label{eq1}
\end{equation}
where $\dot{L}_{wb}$ is the rate of change of angular momentum resulting from magnetized stellar winds according to the prescription of \citet{Matt2015,Matt2019}. As in \citet{Eggenberger2019}, we fixed the constant defining the transition between the unsaturated to the saturated regime to 10.

The quantity $ \dot{L}_{ti}$ expresses the rate of change of the stellar angular momentum  due to the tides induced by the planet in the star:
$$
\dot{L}_{ti} = - \left[\dfrac{1}{2} m_{ pl} \left(\dfrac{\dot{a}}{a} \right)_{ti} \right]\cdot \sqrt{G\left(M_{ st} + m_{ pl} \right)a},
$$
where $m_{ pl}$ is the mass of the planet, $a$ is the orbital distance, $G$ is the gravitational constant and $M_{ st}$ is the mass of the star.
The expression for $(\dot{a}/a)_{ti} $, the secular change in orbit due to dynamical and equilibrium tides, is discussed in Sect.~\ref{sec:2.3}.

Our stellar model will be used to infer the X-ray ($\lambda \sim$ 5-100 \AA) and Ultra-Violet ($\lambda \sim$ 100-920 \AA) luminosity of our star.
Each consists of two contributions: the first is the thermal contribution due to the black-body radiation of the star which can be calculated knowing the total luminosity, by multiplying it with a factor which is a Bose-Einstein integral that comes from integrating the Planck black-body radiation law in the XUV regime. The second is a non-thermal contribution composed of an X-ray and ultra-violet (UV) component, arising due to the radiation emitted by the charged stellar-wind particles which get accelerated by the star's magnetic field.

The non-thermal X-ray luminosity, $L_{ X}$, is computed as in \citet{tu2015}, $L_{ X}=R_{ X} L_{ tot}$, with
$$
R_{ X}=10^{-3.13}{\left(\frac{R_{ o}}{0.196} \right)^{-3.25}},
$$
if  $R_{ o} <  0.196$. For  $R_{ o} \ge 0.196$, $R_{ X}=10^{-3.13}$. $R_{ o}$ is the Rossby number\footnote{The Rossby number is defined here as the ratio between the rotation period of the star, $P_{ rot.}$, and the
convective turn over time, $\tau$, with $\tau={\left(\frac{M_{ env} (R_{ star}-R_{ core})^{2}}{3L_{ star}}\right)^{1/3}}$ taken as in \citep{rasio1996}.}. The expressions above reflect the fact that when the rotation increases ({\it i.e.} $R_{ o}$ decreases), 
then $R_{ X}$ increases and thus, so does the X-ray luminosity.
This increase however, saturates below some critical Rossby number, here equal to 0.196.

The non-thermal UV contribution can be obtained via an empirical relation between the UV and X-ray contributions \citep{sanz2011}.

\begin{figure*}
\centering
\includegraphics[width=.5\textwidth, angle=0]{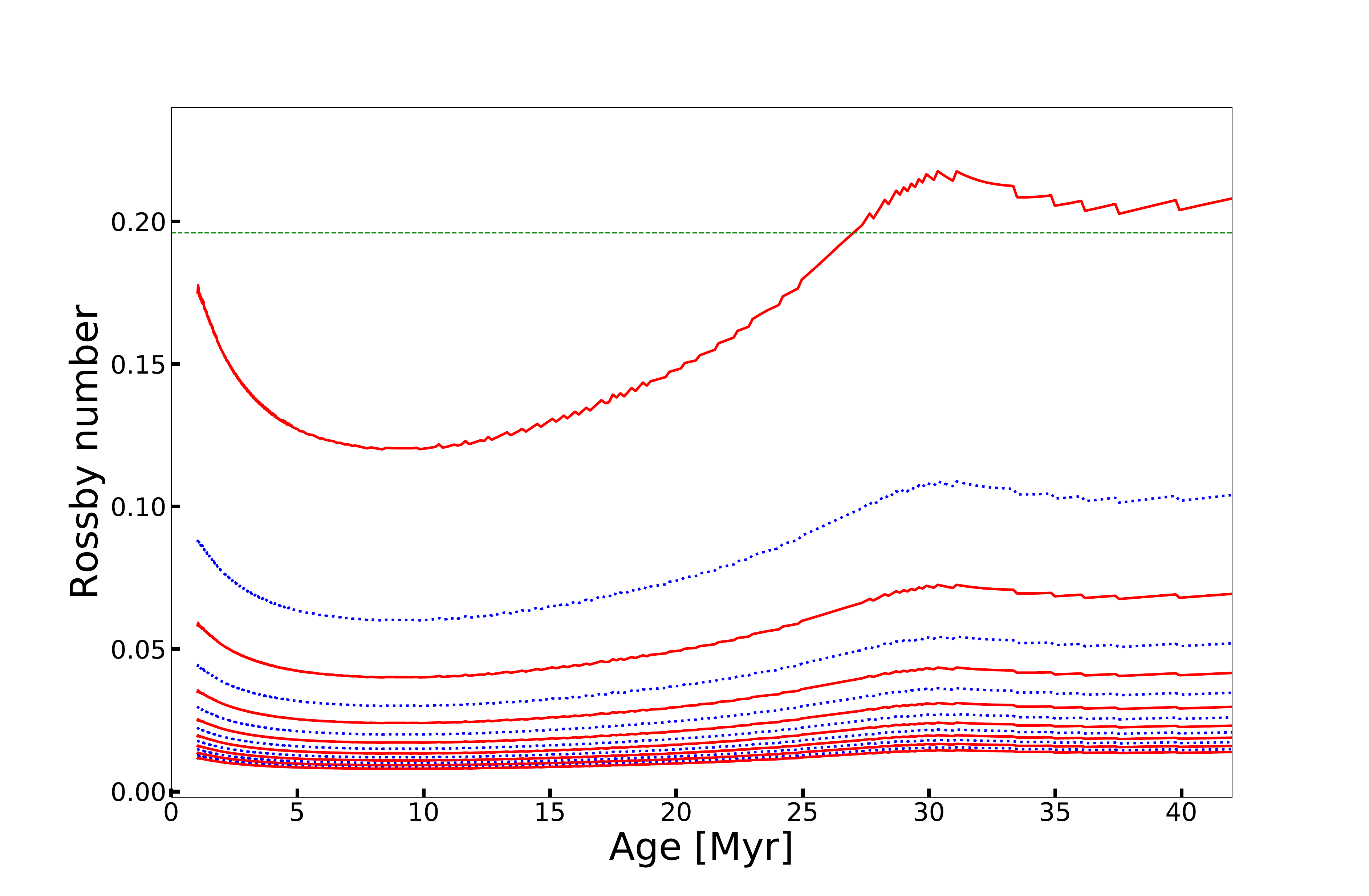}\includegraphics[width=.5\textwidth, angle=0]{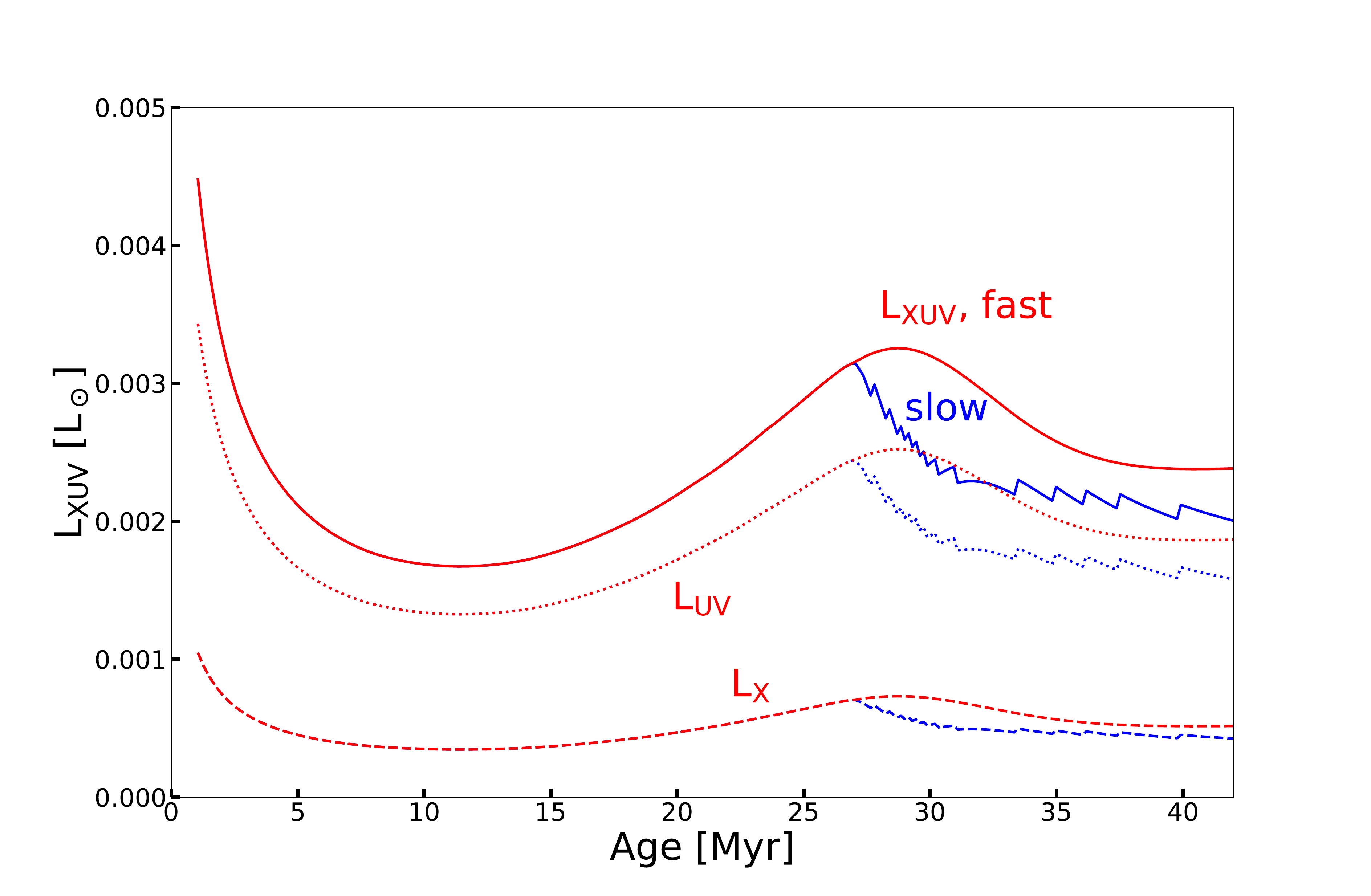}
\caption{{\it Left panel:} Evolution of the Rossby number for an isolated 1 M$_\odot$ stellar model during the Pre-Main-Sequence phase and the very early Main-Sequence phase for different values of the initial rotation. The initial rotation periods and velocities can be seen in Fig.~\ref{ROT_1Msol_ISO}. The Rossby number corresponding to the slowest initial rotation is the upper curve. {\it Right panel:} Evolution of the X-ray, EUV luminosities and of their sum for a slow (in blue) and a fast (in red) rotating 1 M$_\odot$ model. The slow and fast rotation correspond respectively to the slowest and fastest model shown in  Fig.~\ref{ROT_1Msol_ISO}. 
}
\label{Rossby}
\end{figure*}

In the left panel of Fig.~\ref{Rossby}, the evolution of the Rossby number is shown as a function of time for 1 M$_\odot$ stellar models with different initial rotations. We see that for most of the present models, the Rossby number is below 0.196. This means that the X-ray luminosity will follow the total luminosity scaled with a constant factor equal to $0.00074$. The luminosities in X-rays, Ultra-Violet and their sum is shown in the right panel of Fig.~\ref{Rossby}. Saturation is always reached for the fast rotating model, while for the slow rotating one, it occurs only for ages below about 27 Myr. During the saturation period, the red and blue curves superpose, so only the red curve is shown in Fig.~\ref{Rossby}.

\subsection{Planet mass-evaporation rates}

The evaporation rate is taken as in \citet{erkaev2007}:
\begin{equation}
 \dot{m}_{ pl} ={ \frac{\epsilon \pi\ (\beta r_{ pl})^{2} F_{ XUV}}{\Delta \Phi}
 },
 \label{eq2}
\end{equation}
where $\epsilon$ 
is the heating efficiency. This quantity, $\epsilon$, is smaller than unity because 
part of the energy can be used for ionisation. Also processes like recombinations, collisional excitations etc. can cause radiative cooling that will reduce the energy available for the heating.
Here we assume $\epsilon=0.4$, as suggested by \citet{valencia2010}. These authors showed that this value is adapted to the cases of hot Jupiters and strongly
irradiated rocky planets. $\beta$ is a number larger than unity,
accounting for the inflation of the envelope due to its heating by the irradiation of the star. $\beta$ is taken as in \citet{Salz2016}. $r_{ pl}$ is the planet radius.
We assume a spherical planet of uniform and constant density. The initial planet radius is computed from the initial mass of the planet using the mass-radius relation found by \citet{bashi2017}. This relation is a piece-wise power law relation 
obtained by using advanced statistical techniques to fit the power law with 
direct observational exoplanet data and with numerical simulations of planet formation (see Fig.~\ref{Rpl}). $F_{ XUV}$ is given by
$L_{ XUV}/(4\pi a^2)$, where $a$ is the orbital distance of the assumed
quasi-circular orbit of the planet. $\Delta \Phi$ is the energy required to lift one unit of mass up to the Roche lobe and is taken as in
\citet{erkaev2007}. We applied this evaporation law for planet masses in the range 6 up to 800 Earth masses (0.02 and 2.5 Jupiter masses) as in \citet{Salz2016}. We note that under the present evaporation law, planets more massive than 2.5 M$_{ Jup}$ would lose only a negligible amount of mass and thus, their evolution would correspond to a constant mass evolution.

The change of the planet radius (under the assumption of constant bulk density of the interior structure in our simple planet model) is obtained from 
\begin{equation}
 \dot{r}_{ pl} = \dfrac{1}{3}\dfrac{\dot{m}_{ pl}}{m_{ pl}} r_{ pl}.
 \label{eq3}
\end{equation}

\subsection{Orbital evolution}\label{sec:2.3}

We consider here the case of a planet orbiting in a quasi-circular orbit around its host star. Secular change of the orbital distance due to equilibrium and dynamical tides produced by the planet on the star is considered. Tides are accounted for only when convective zones are
present in the star. Note that since we are considering here the case of a 1 M$_\odot$ star, an outer convective envelope is present during the whole evolution. We also account for the secular change of the orbital distance due to change in planet mass from mass-evaporation. The expression for the net secular change in the planetary orbit is here written,
\begin{equation}
    \left(\frac{\dot{a}}{a}\right) = -\frac{\dot{m}_{pl}}{m_{pl}+M_{st}} + \left(\frac{\dot{a}}{a}\right)_{eq} + \left(\frac{\dot{a}}{a}\right)_{dy},
    \label{eq4}
\end{equation}
where the first term is due to planet mass-evaporation and the last two terms, being the contributions due to the equilibrium tides and dynamical tides respectively, together are equal to the quantity $(\dot{a}/a)_{ti} $ mentioned in Sect.~\ref{sec:2.1}. All these quantities are computed as in \citet{rao2018}\footnote{In \citet{rao2018}, we
showed that the expression of the equilibrium tide, $(\dot{a}/a)_{eq}$, can be expressed
using a quantity $\sigma_\star$. This quantity is sometimes kept constant by some authors while, as shown in Fig.~1 of \citet{rao2018}, it actually varies as a function of time, especially during the PMS phase and the red giant phase. We
note here an error in the caption of this Fig.~1. The quantity plotted is 36 times the expression given in Eq.~4 of \citet{rao2018} and not directly the expression given in Eq.~4. This does not change, however, the main point about the variation of this quantity with time.}. 

In the present work, we neglect changes of the stellar mass due to stellar winds, 
and therefore, the planet drag and the planet mass accretion.

\subsection{Numerical procedure}

In comparison to our previous work \citep{rao2018}, the present version of the code differs mainly by the following points: We account for the planet evaporation process. We consider here that the star rotates as a solid body during the PMS phase and MS phase. The magnetic braking law has also been updated (see Sect.~\ref{sec:2}). The time-step variations have been improved to allow the code to evaluate more accurately the evolution of very close-in planets. The present code
solves a system of four coupled ordinary differential equations, for the stellar angular momentum, the mass and radius of the planet, and the orbital distance, i.e. equations \eqref{eq1}, \eqref{eq2}, \eqref{eq3} and \eqref{eq4} respectively. Since time does not explicitly occur in any of these equations, we use the explicit classic fourth-order Runge-Kutta method for solving this system of equations.

\section{Photoevaporation and tides}\label{sec:3}

\subsection{Consistency check of the new version of our code}

\begin{figure}[h]
\centering
\includegraphics[width=.5\textwidth, angle=0]{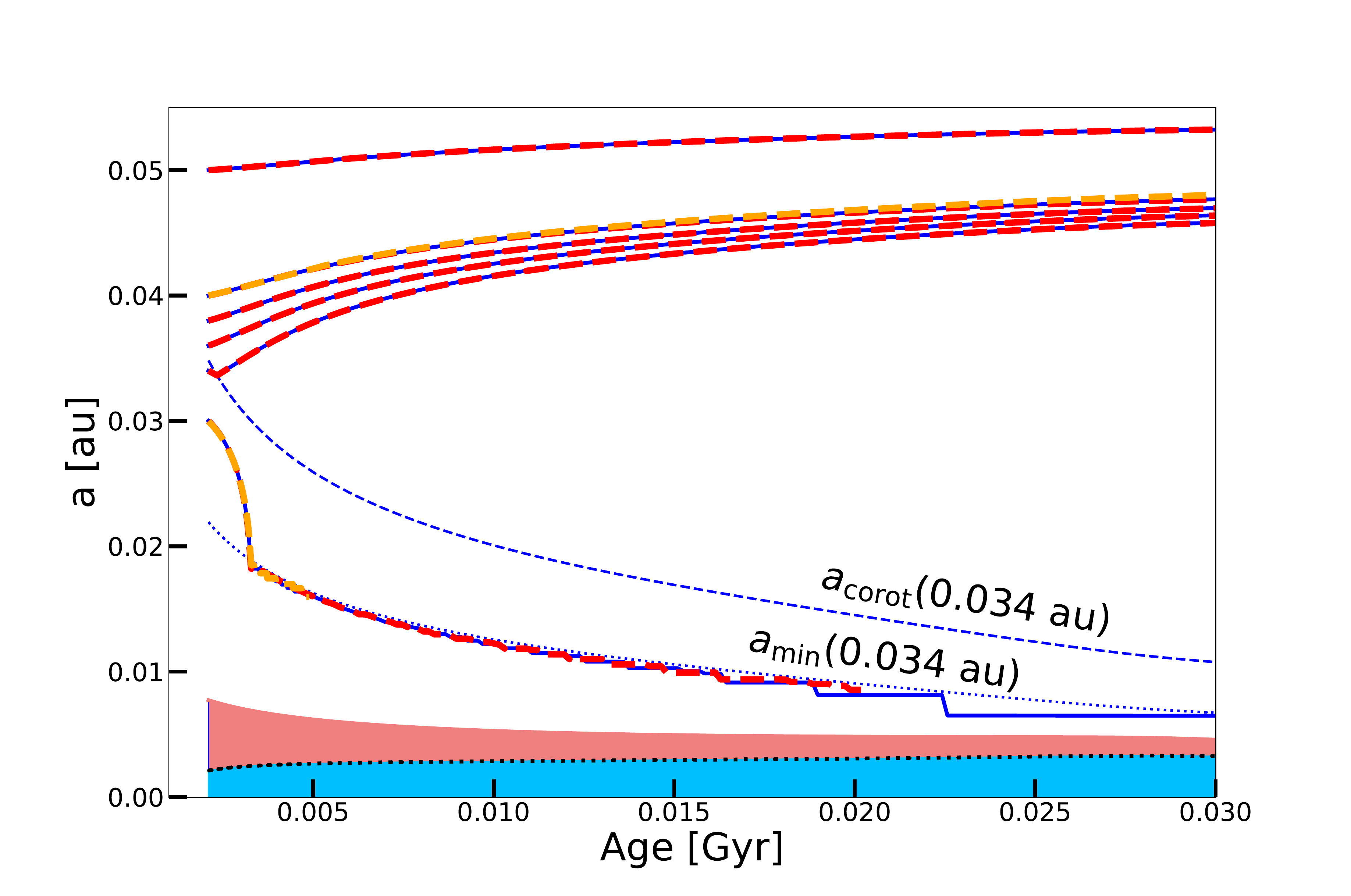}
\caption{Evolution as a function of time of the orbits of 1 $M_{Jup}$ mass planets around a 1 M$_\odot$ star as obtained using different approaches.
The blue continuous lines are obtained in the present work by switching-off the planet evaporation process. The red dashed lines are obtained switching-on the evaporation process.
The orange dashed lines (only two cases shown) are the orbits obtained in \citet{rao2018}. 
The blue dashed and dotted lines show respectively, the corotation radius ($a_{ corot}$) and the critical distance below which dynamical tides are no longer active ($a_{ min}$). They are obtained for the orbit beginning its evolution at a distance equal to 0.034 au. The upper-limit of the pink zone corresponds to the radius of the star in au. Its lower limit shows the radius at the base of the stellar convective envelope. The blue zone indicates the size of the stellar radiative interior.}
\label{figlinkSI}
\end{figure}

In order to see how the results obtained with the new version of our code compare to the results presented in \citet{rao2018}, we plot a few orbits obtained in \citet{rao2018} for a 1 M$_{ Jup}$ mass planet around a 1 M$_\odot$ star together with those computed with the present version in Fig.~\ref{figlinkSI}.

Comparing the dashed orange lines \citep[models by ][]{rao2018}, and the continuous lines (current models without evaporation), we see only very small differences linked to the models used for the stellar rotation \citep[non-solid body rotation in ][and solid body rotation here]{rao2018} and also due to the different numerical procedure used in the present version. Concerning the latter point, we see that the orange curve for the planet 
beginning its evolution at 0.03 au 
stops at a much earlier stage than the blue one. This is because in the old version, due to a less sophisticated way of handling the variation of the time steps, the orbit plunged very early into the star. Having known this reason, as already explained in \citet{rao2018}, we did not show that part of the orbit \citep[see Fig.~2 in][]{rao2018}. In the new version, this point has been improved and that is why, the orbit can be computed on a longer timescale. Earlier, we were unsure of the fate of the planet corresponding to this computation, but the present calculation shows that in the absence of any evaporation process, the planet would be engulfed after a few 100 Myr. 

From the discussion above, we conclude that there is
consistency between the results we obtained in \citet{rao2018}
using the old version of the code and the present results obtained using the new version.

\begin{figure*}
\centering
\includegraphics[width=.25\textwidth, angle=0]{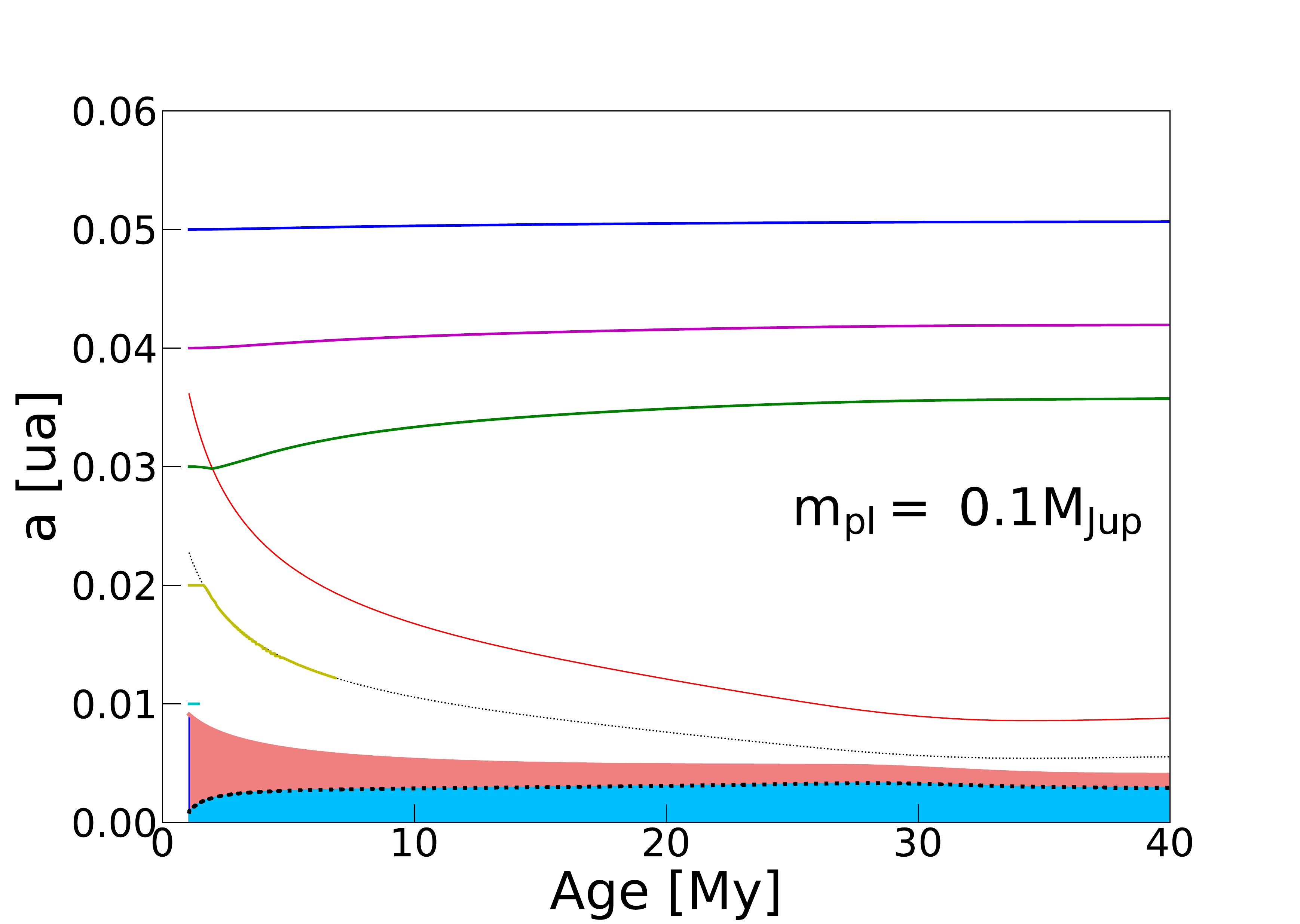}\includegraphics[width=.25\textwidth, angle=0]{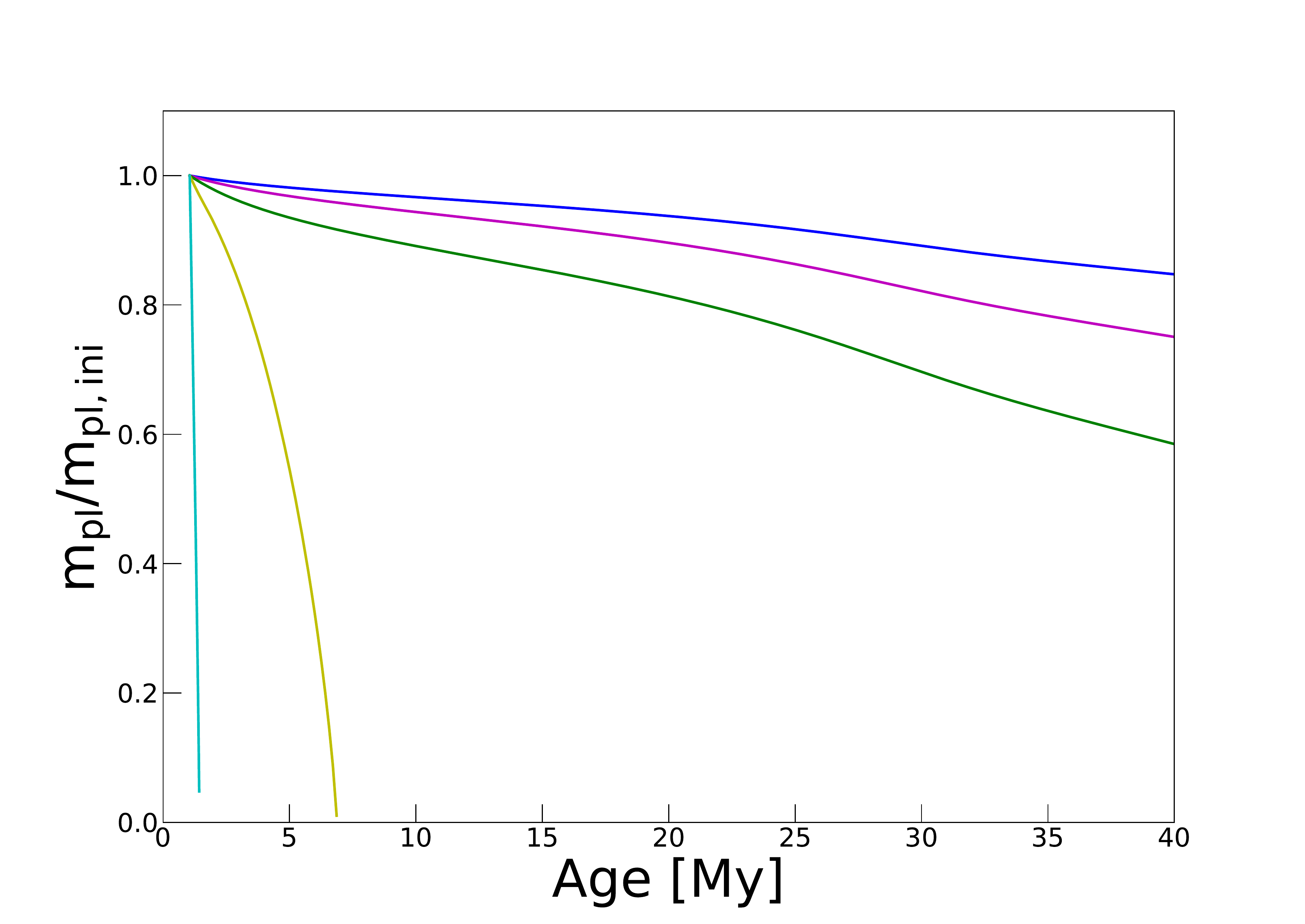}\includegraphics[width=.25\textwidth, angle=0]{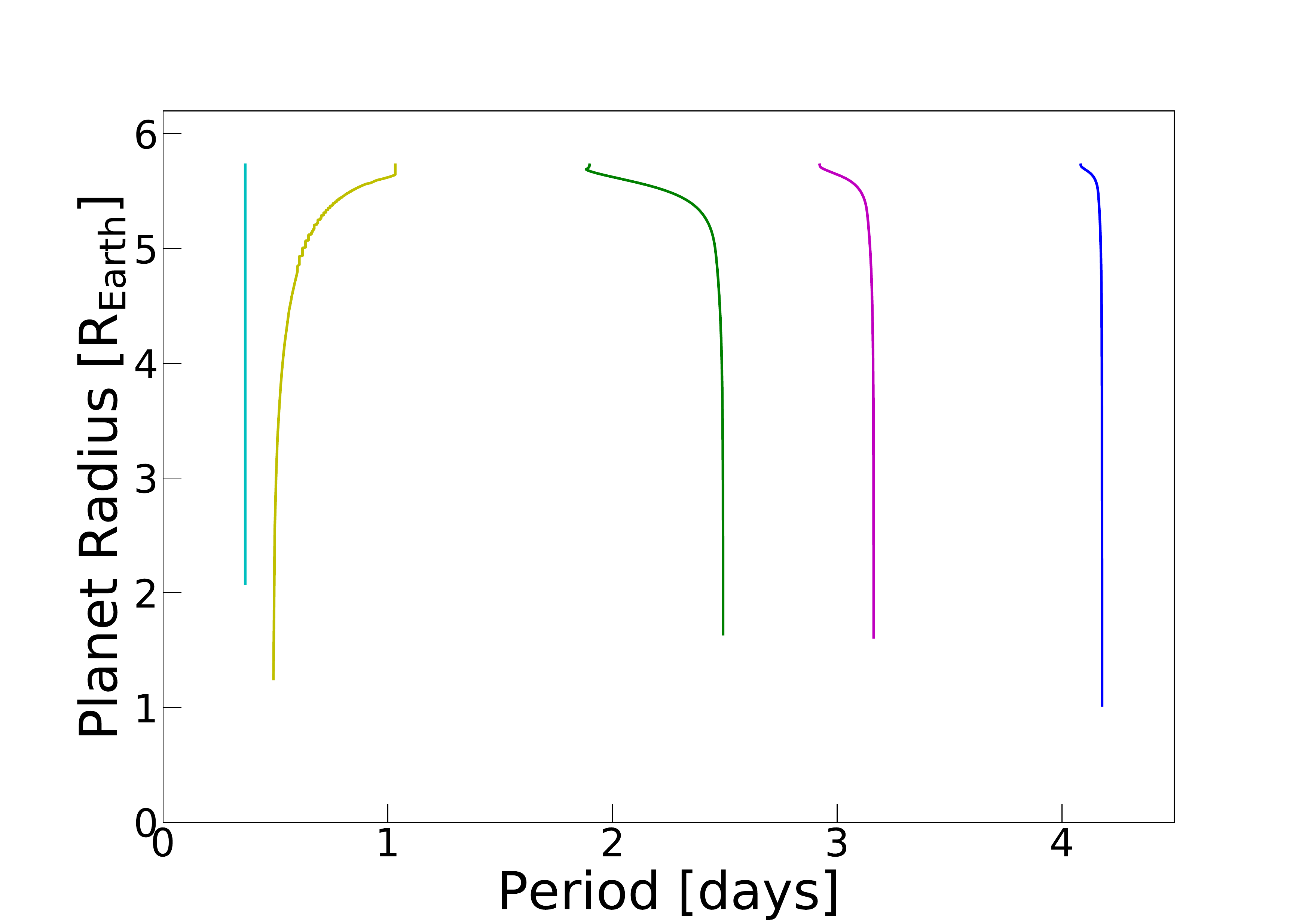}\includegraphics[width=.25\textwidth, angle=0]{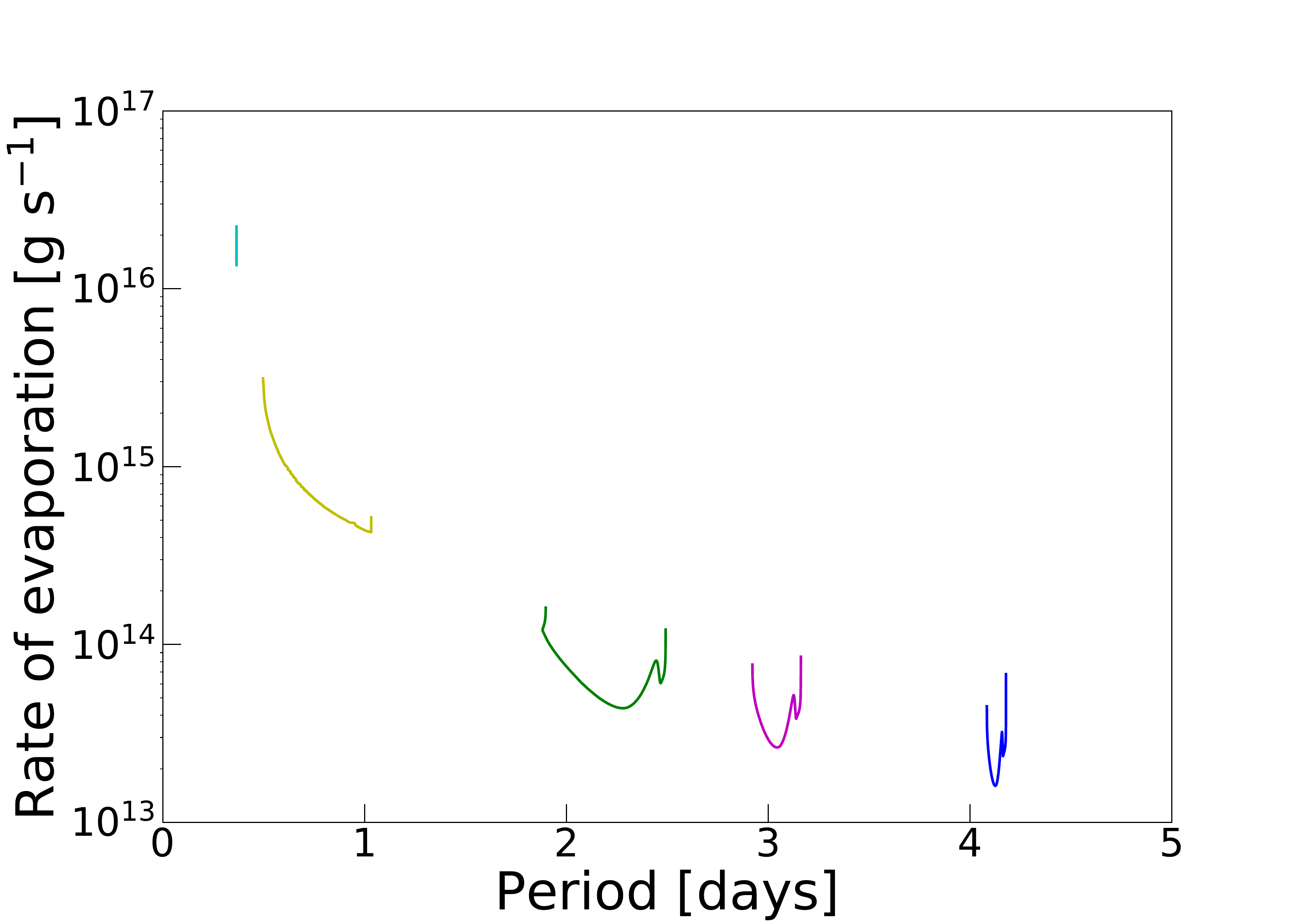}
\includegraphics[width=.25\textwidth, angle=0]{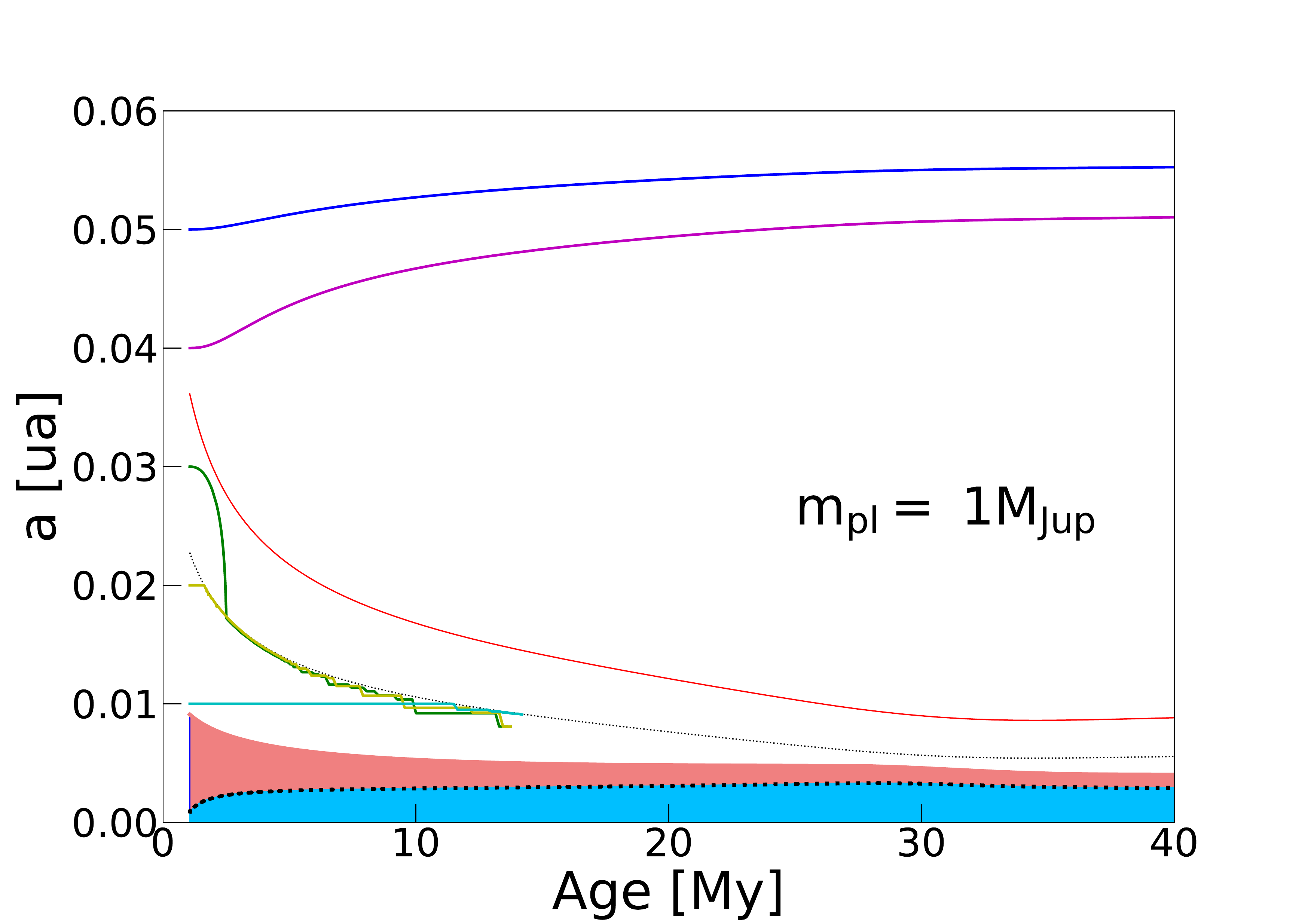}\includegraphics[width=.25\textwidth, angle=0]{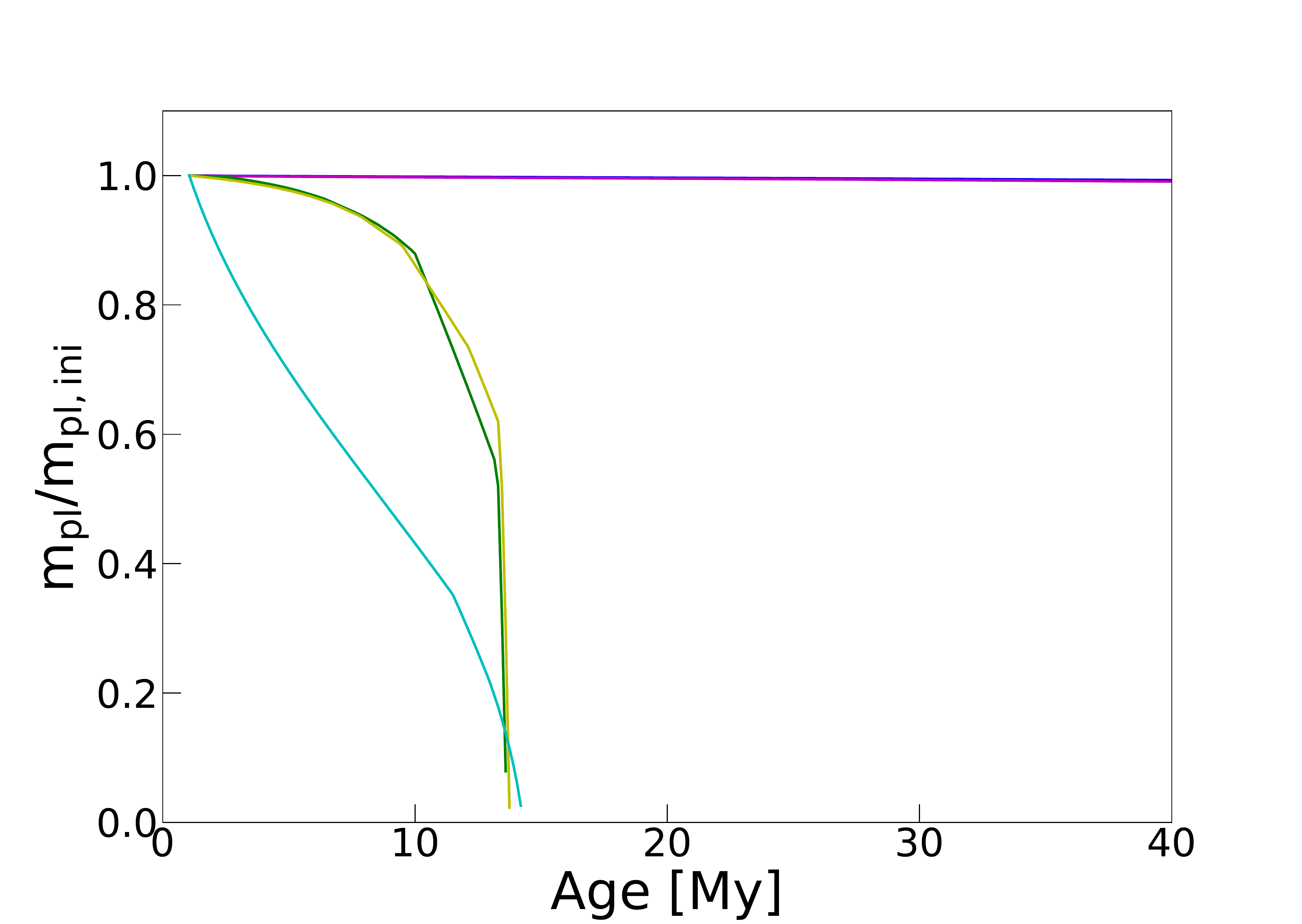}\includegraphics[width=.25\textwidth, angle=0]{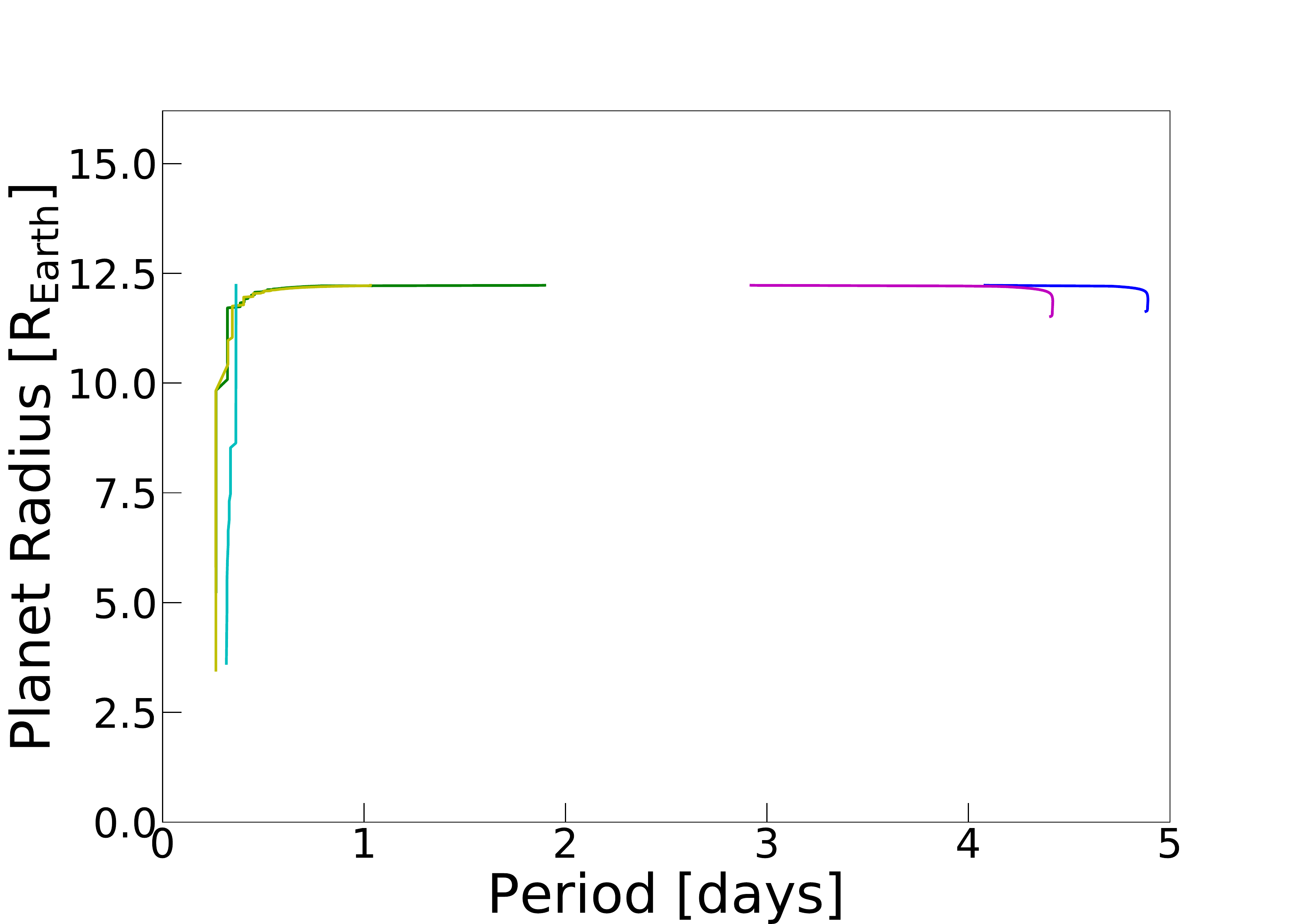}\includegraphics[width=.25\textwidth, angle=0]{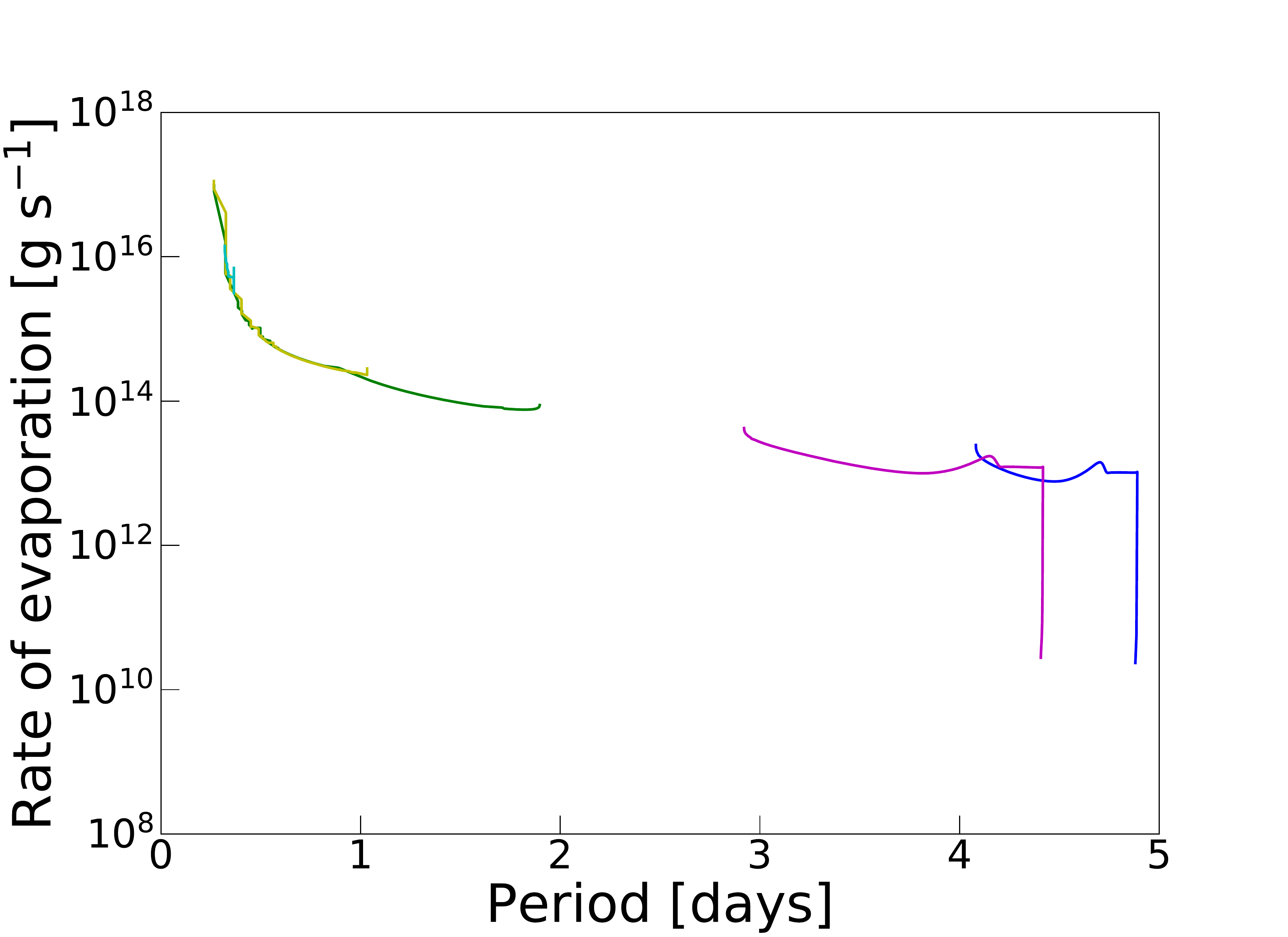}
\includegraphics[width=.25\textwidth, angle=0]{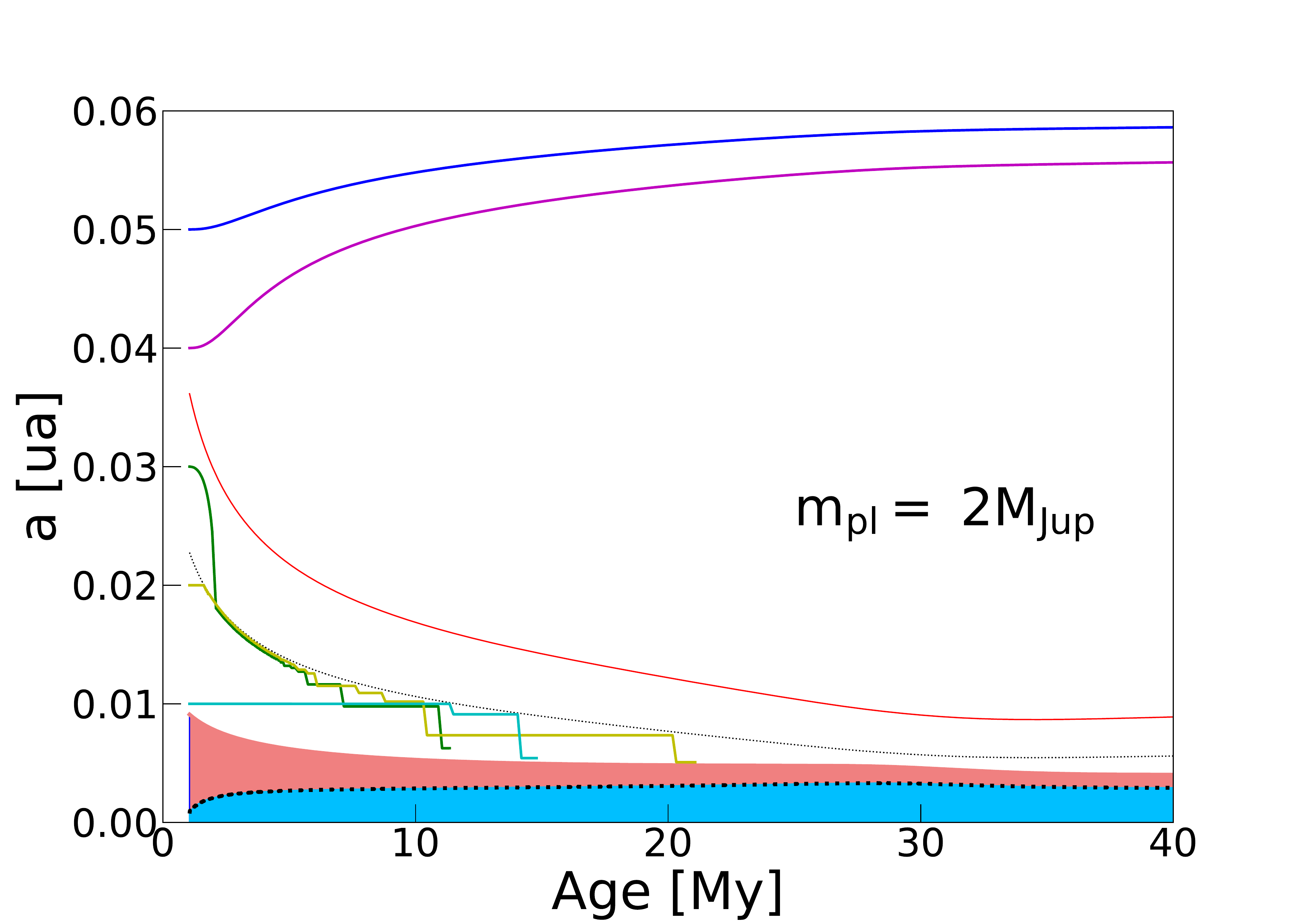}\includegraphics[width=.25\textwidth, angle=0]{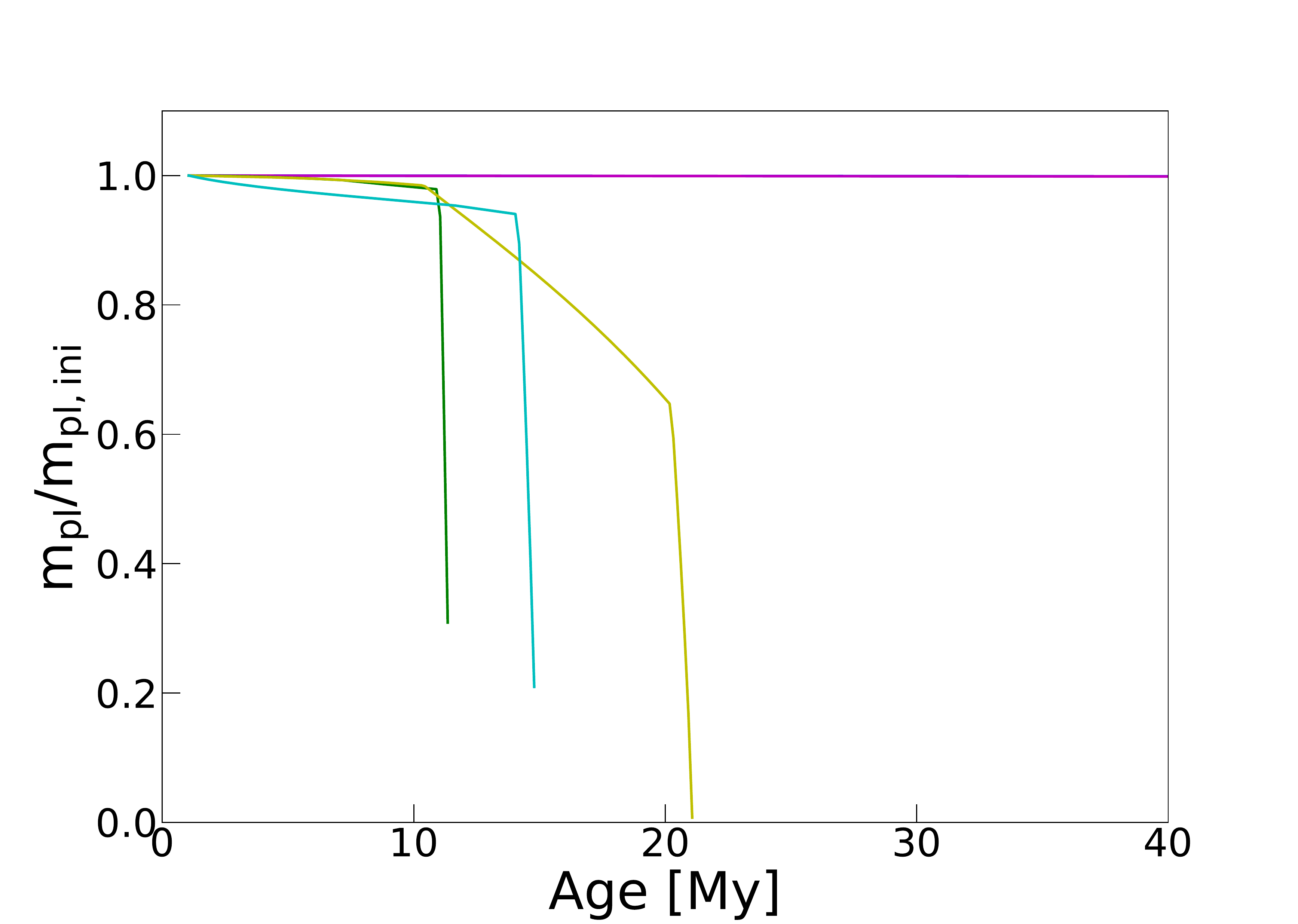}\includegraphics[width=.25\textwidth, angle=0]{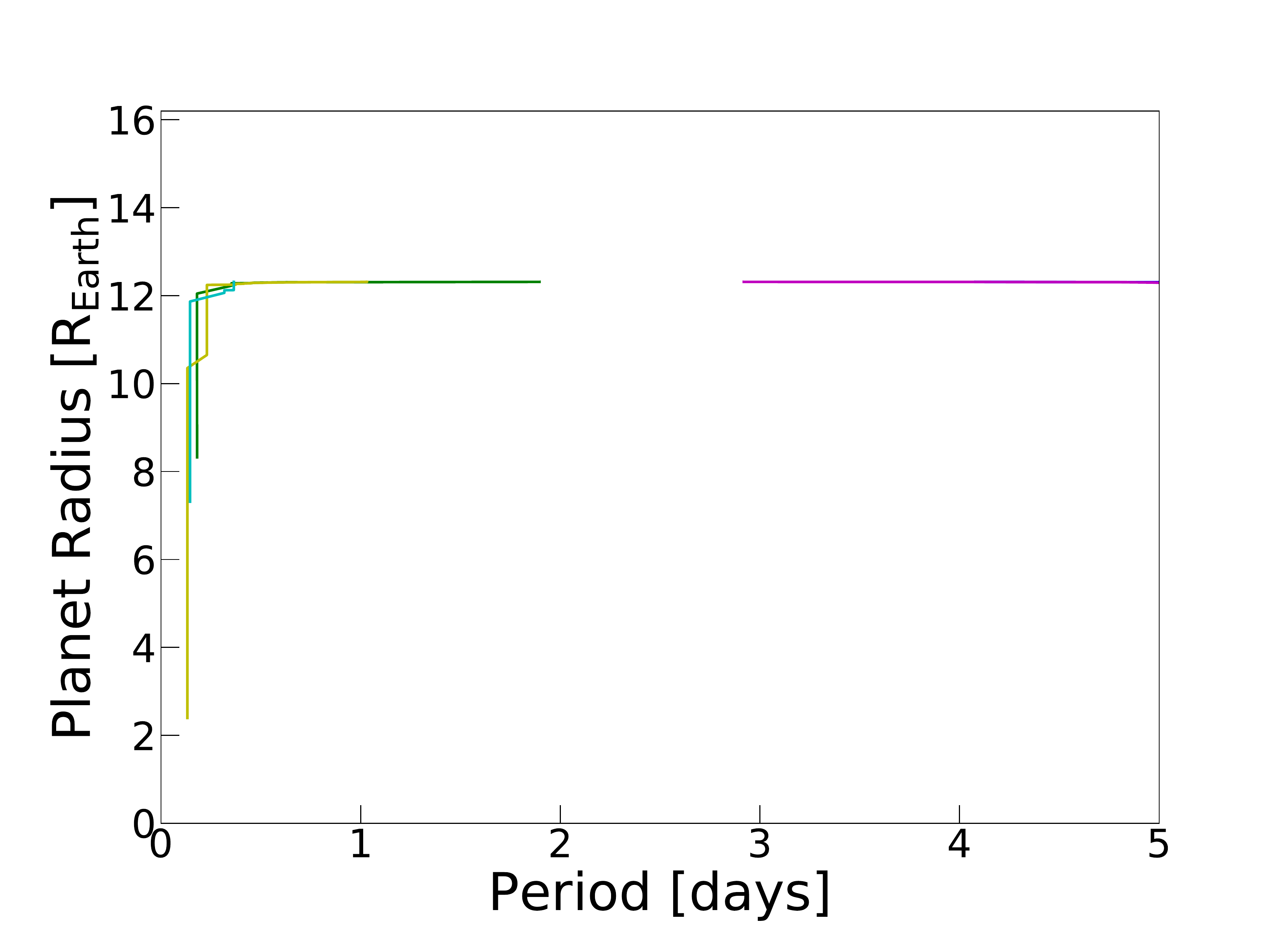}\includegraphics[width=.25\textwidth, angle=0]{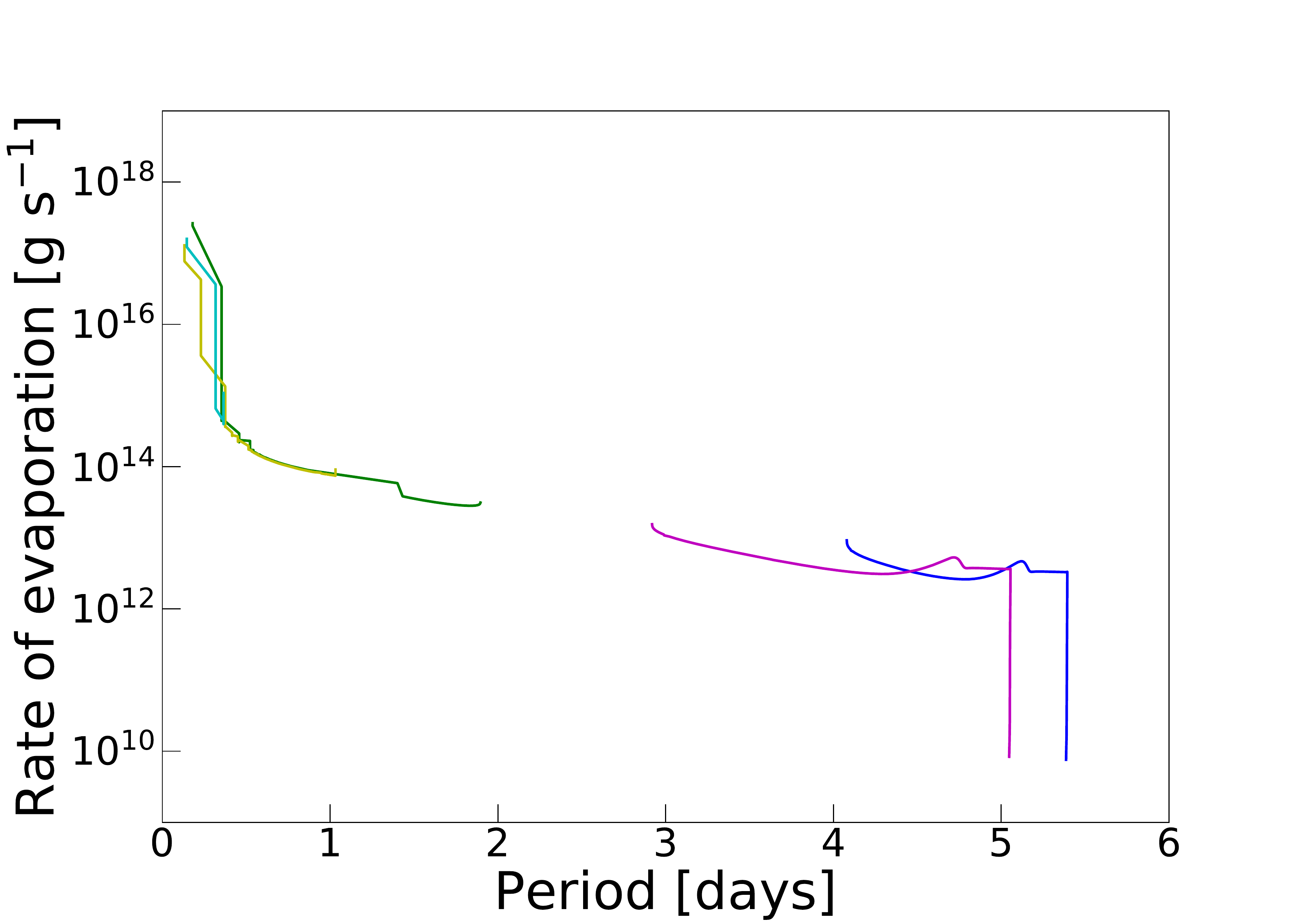}
\includegraphics[width=.25\textwidth, angle=0]{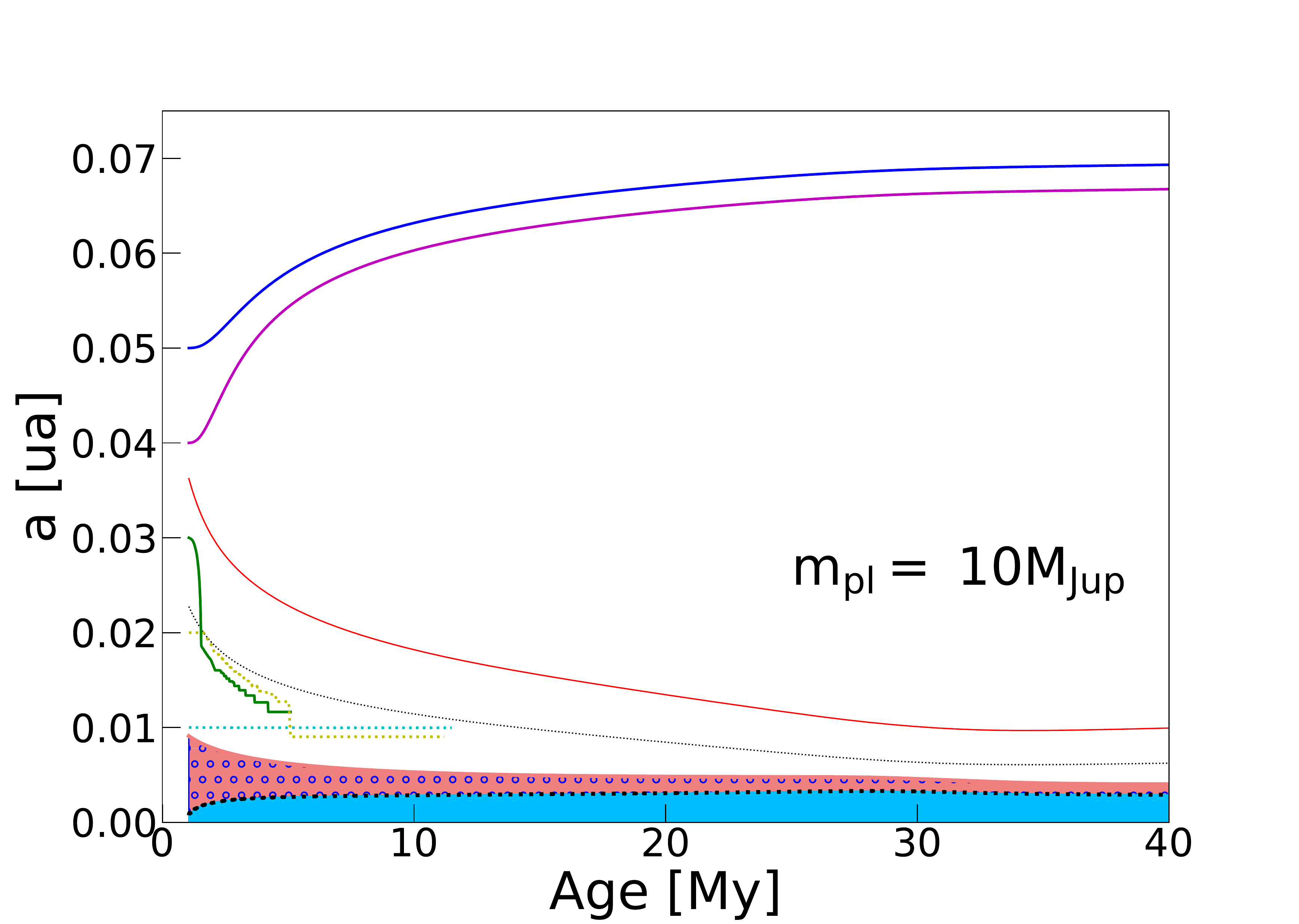}\includegraphics[width=.25\textwidth, angle=0]{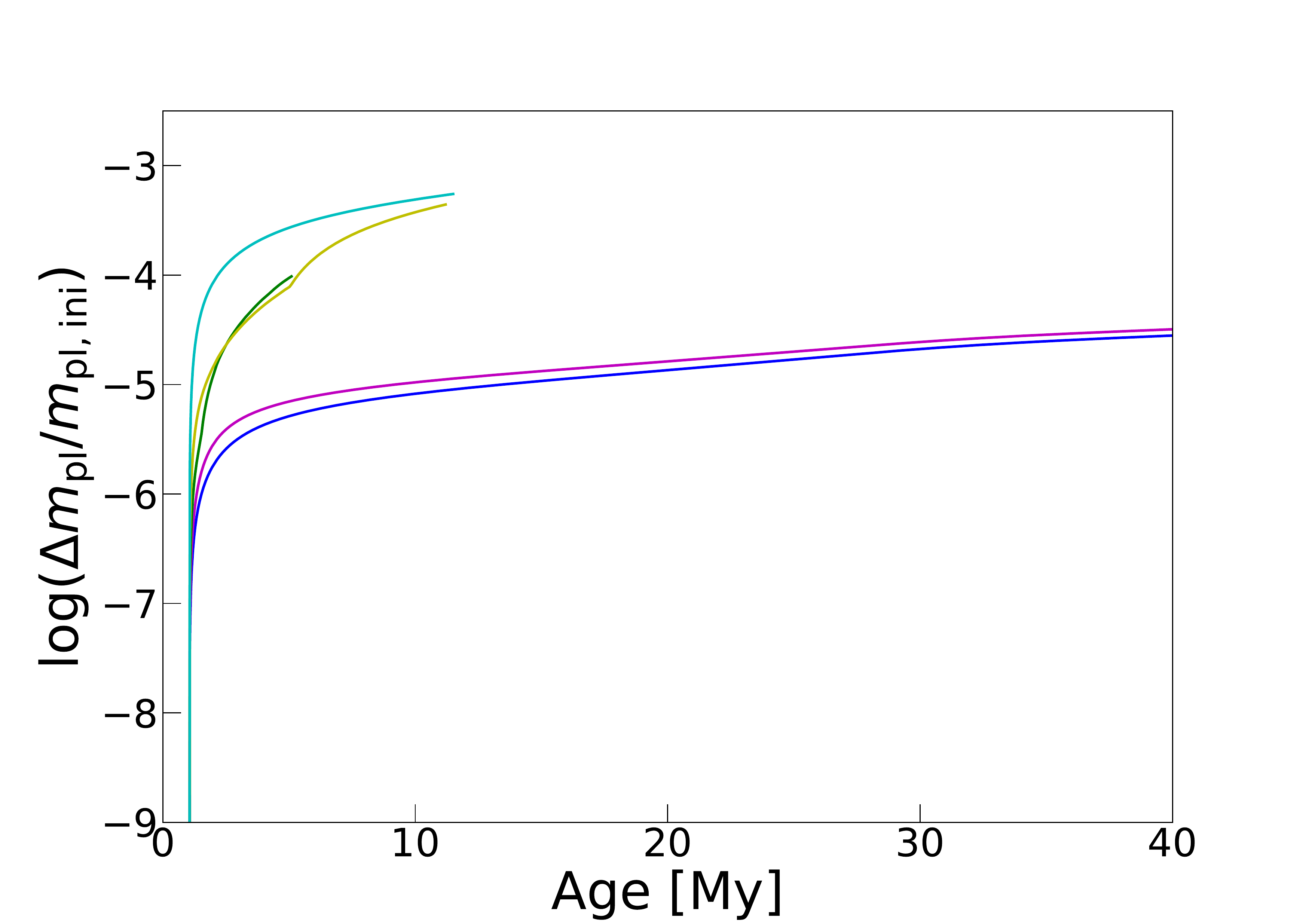}\includegraphics[width=.25\textwidth, angle=0]{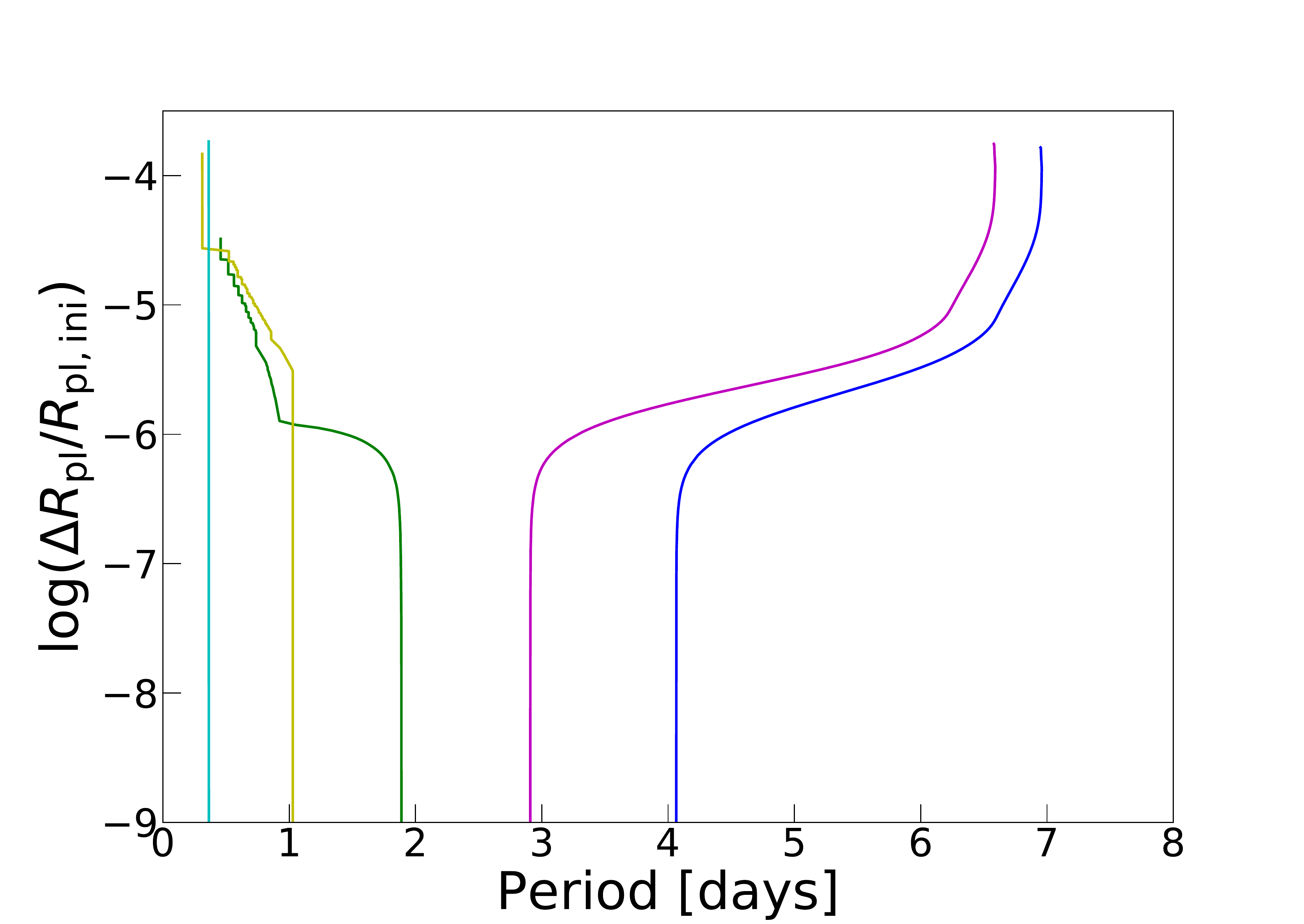}\includegraphics[width=.25\textwidth, angle=0]{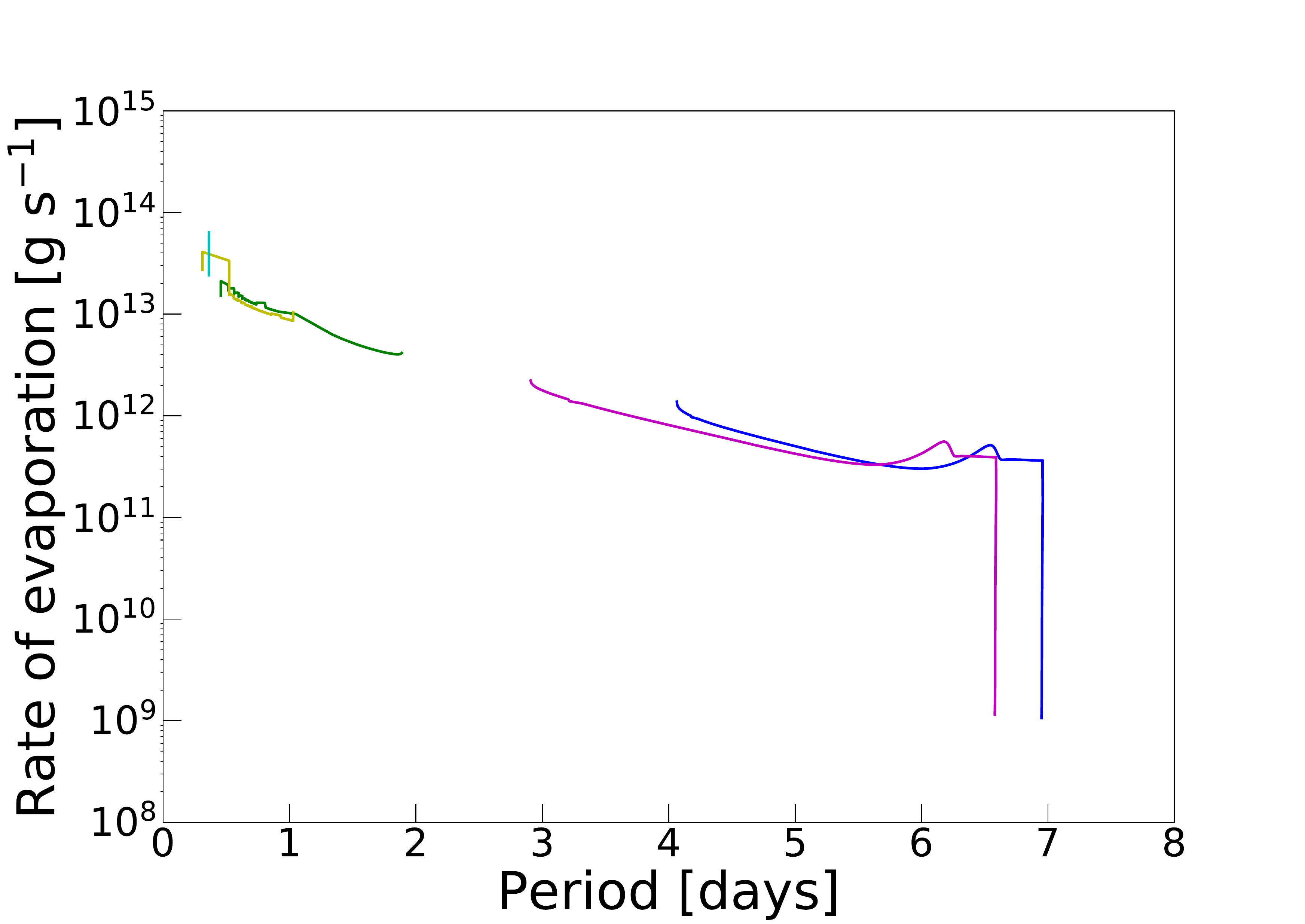}
\caption{Each row shows the evolution, for a given planet mass and for different initial radii of the orbit, of the orbital distance as a function of time (first column panel), of the planet mass as a function of time (second column), of the
planet radius as a function of the orbital period (third column), and of the evaporation rate as a function of orbital period (fourth column). Identical colors in a set of row panels correspond to models with the same initial distance to the star (see the left panel in each raw). Note that panels of the fourth row in the second and third column have a different ordinate.}
\label{figOrMa}
\end{figure*}

\subsection{Impact of evaporation and tides on a Jupiter mass planet}

The magenta lines in Fig.~\ref{figlinkSI} shows the evolution of the orbits of a Jupiter mass planet when the evaporation processes are accounted for. At first sight, there are not many differences between the evolution of the planet orbits  with and without evaporation. However, as shown below (see the panel in the second column, second row of Fig.~\ref{figOrMa}), depending on the initial distance of the planet to the star, we can have a significant change of the mass of the planet for distances below 0.04 au. As a numerical example, we find that the planet, having begun its evolution at a distance of 0.03 au, would be obliterated at an age of 20 Myr due to evaporation and not be engulfed at 100 Myr as predicted by non-evaporating model. We assume here a completely gaseous planet, which is not very realistic, but we may interpret that the photo-evaporation process will be efficient enough for completely removing the planet's atmosphere at least. Afterwards, the remaining rocky core of the planet may pursue its evolution. Evaporation may have thus a strong impact for Jupiter mass planets whose orbits shrink as a result of the dynamical tides.
The present simulations indicate that all planets of this mass beginning their evolution at a distance less than 0.034 au are evaporated during the Pre-Main-Sequence phase (this number is valid for the considered initial rotation of our 1 M$_\odot$ star model).

\begin{figure*}[h]
\centering
\includegraphics[width=.5\textwidth, angle=0]{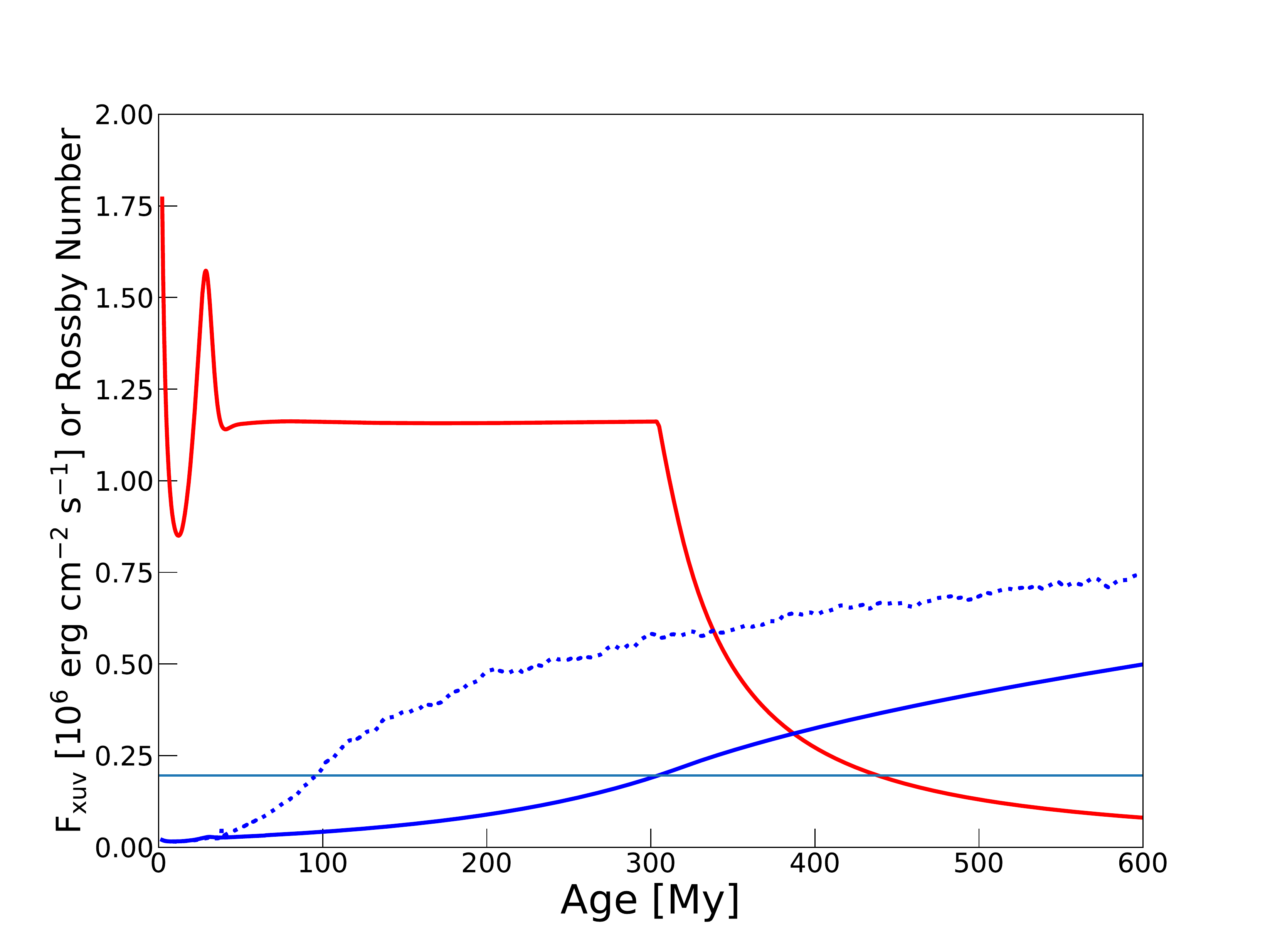}\includegraphics[width=.53\textwidth, angle=0]{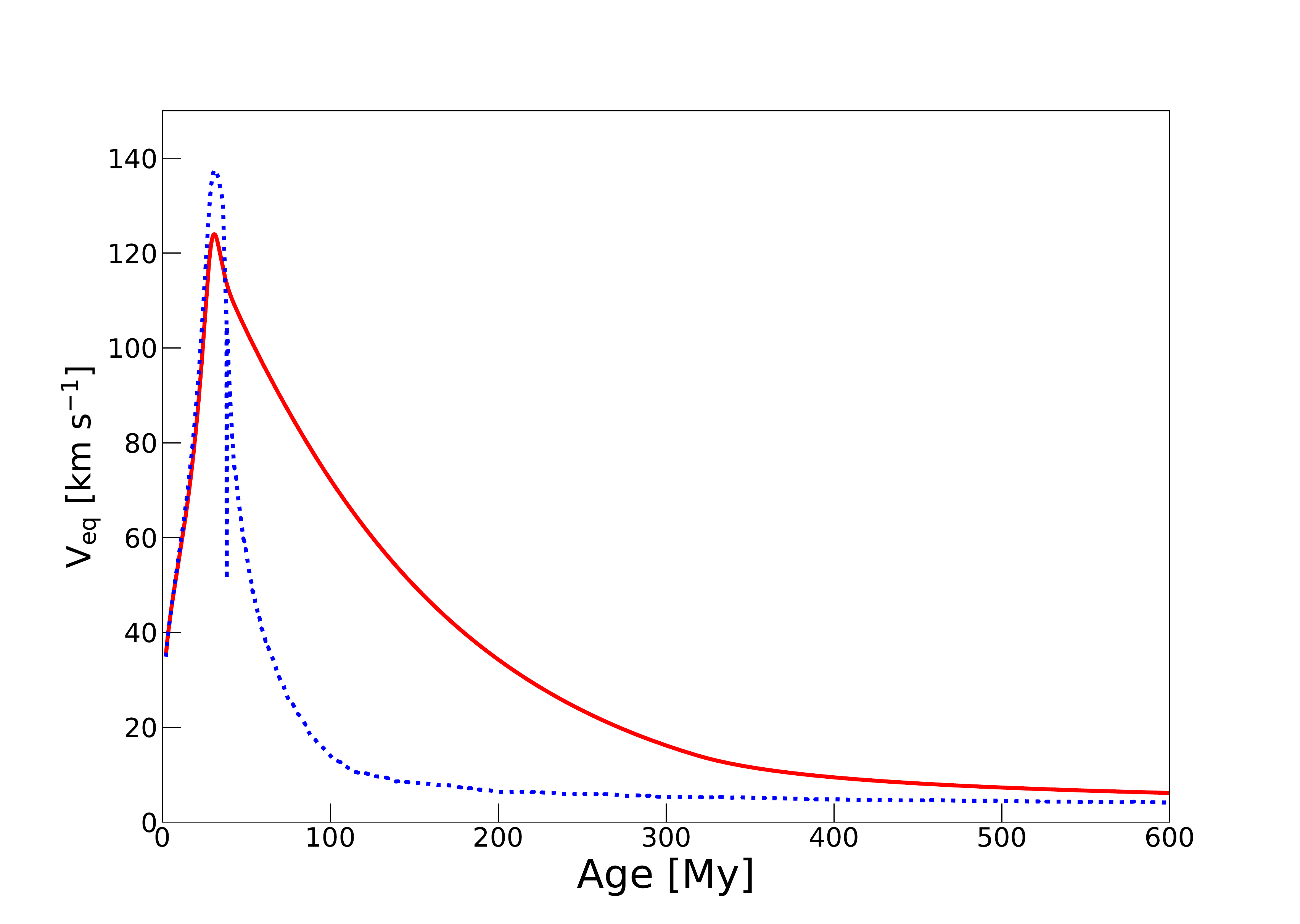}
\caption{
{\it Left panel:} The red continuous line shows the evolution as a function of time of the XUV flux on the planet beginning its evolution at a distance of 0.05 au. The blue continuous line shows the evolution
of the Rossby number obtained from the solid body stellar rotation model. The dotted blue line is the Rossby number obtained from the differentially rotating model. Below the horizontal line at 0.196, the X-ray luminosity is a constant fraction (0.00074) of the bolometric luminosity. When the curve is above the horizontal line, the X-ray luminosity decreases rapidly as the Rossby number increases (see text).
{\it Right panel:} Evolution of the surface velocity of the 1 M$_\odot$ model. The red line shows the surface velocity assuming solid body rotation. The blue line is the surface velocity obtained from the model used in \citet{rao2018} where differential rotation occurs due to the internal angular momentum redistribution resulting from shear instability and meridional currents.
}
\label{figVFR}
\end{figure*}

The evaporation process can be accelerated by the action of the dynamical tides that on one hand shrinks the orbit and thus imposes a stronger high-energy flux received by the planet, and also by the fact that the angular momentum lost by the orbit of the planet is transferred to the star, which therefore rotates faster and has an increased radiation activity. The latter point, however, does not play a significant role here because during this phase, the X-ray and UV luminosities are independent of the stellar rotation, because they are saturated (see Fig.~\ref{figVFR}). However the duration of the phase during which saturation occurs depends on the initial rotation. This duration is around 170 and 570 Myr if the initial rotation is respectively equal to 3.5 and 70 km s$^{-1}$.

In case saturation was not achieved, let us investigate whether the change of the orbital angular momentum is sufficient to significantly spin-up the star. The spin angular momentum of the star is of the order $\sim  1.65 \times 10^{50}$ g cm$^2$ s$^{-1}$, while the initial orbital angular momentum of the planet is less than 10\% of this value. Therefore, even if the totality of the orbital angular momentum were to be transferred to the star, it would produce an increase of the surface velocity by around 10\% (if the stellar structure is kept constant). In reality, only a fraction of the initial orbital angular momentum can be transferred to the star, since in our computation, the planet is completely evaporated at some distance from the host star. This causes the surface velocity of the star (at the time when the planet disappears due to the evaporation process) to be only 5\% larger than the velocity it would have had in the absence of any planet. Outside the saturation regime, such a stellar spin-up would cause an increase of the XUV flux by about 15\%, which would be non-negligible. 

For the cases where the planets survive, we find that photo-evaporation decreases their total mass by a fraction between 11 and 16\% at the end of the MS phase for the initial distances equal to 0.05 and 0.04 au respectively.
Most of the mass loss will occur during the first 300 Myr, when the Rossby number is below 0.196 (see the blue continuous line in the left panel of Fig.~\ref{figVFR}). As soon as the Rossby number becomes larger than this limit, one sees that the XUV flux decreases rapidly. This occurs, in our model, when the surface velocity becomes inferior to about 16 km s$^{-1}$ (see the right panel of Fig.~\ref{figVFR}). We note that, starting with the same angular momentum content, a model in which differential rotation occurs shows a very different evolution for the surface velocity than a model rotating as a solid body. The differences are small during the contracting phase towards the Zero Age Main-Sequence (the ZAMS corresponds to the peak, occurring around 30 Myr), but are important during the MS phase. The solid body rotating model
shows larger surface velocities\footnote{In the differentially rotating model, there is no instantaneous coupling between the envelope and the core. This implies that the envelope is slowed down by the wind magnetic braking first and it takes some time for this braking to be communicated to deeper layers, or for the deeper layers to mitigate the slowing down of the surface. Thus, in the differentially rotating models, the braking of the surface is much more efficient.} that keep the Rossby number below 0.196 for a much longer period (compare the continuous blue line for the solid body rotating model with the dotted one for the differentially rotating one). This illustrates well the fact that, starting from the same initial angular momentum content, depending on the way angular momentum is transported inside the star, very different results can be obtained.

\subsection{Other planet masses}

Fig.~\ref{figOrMa} presents time-evolutions of the planet orbit and planet mass during the first 40 Myr, as well as the variations of the planet radius and of the evaporation rate as a function of the orbital period.
We considered here a rotation of the star equal to about 40 km s$^{-1}$, 40 Myr before the star reaches the ZAMS. 

As has been obtained by many previous works, evaporation will
impact mostly low mass planets \citep[see e.g. references in the recent review by][]{Owen2019}.
From the present simulations, we distinguish  two
planet-mass ranges: the intermediate and high-mass planets showing different behaviours under the action of tides and evaporation\footnote{As already mentioned earlier, in the following when we say that a planet is completely evaporated, we mean that this would be the case for the kind of planet structure considered in the present work (see Sect.~2.2).}:

\begin{figure}[h]
\centering
\includegraphics[width=.5\textwidth, angle=0]{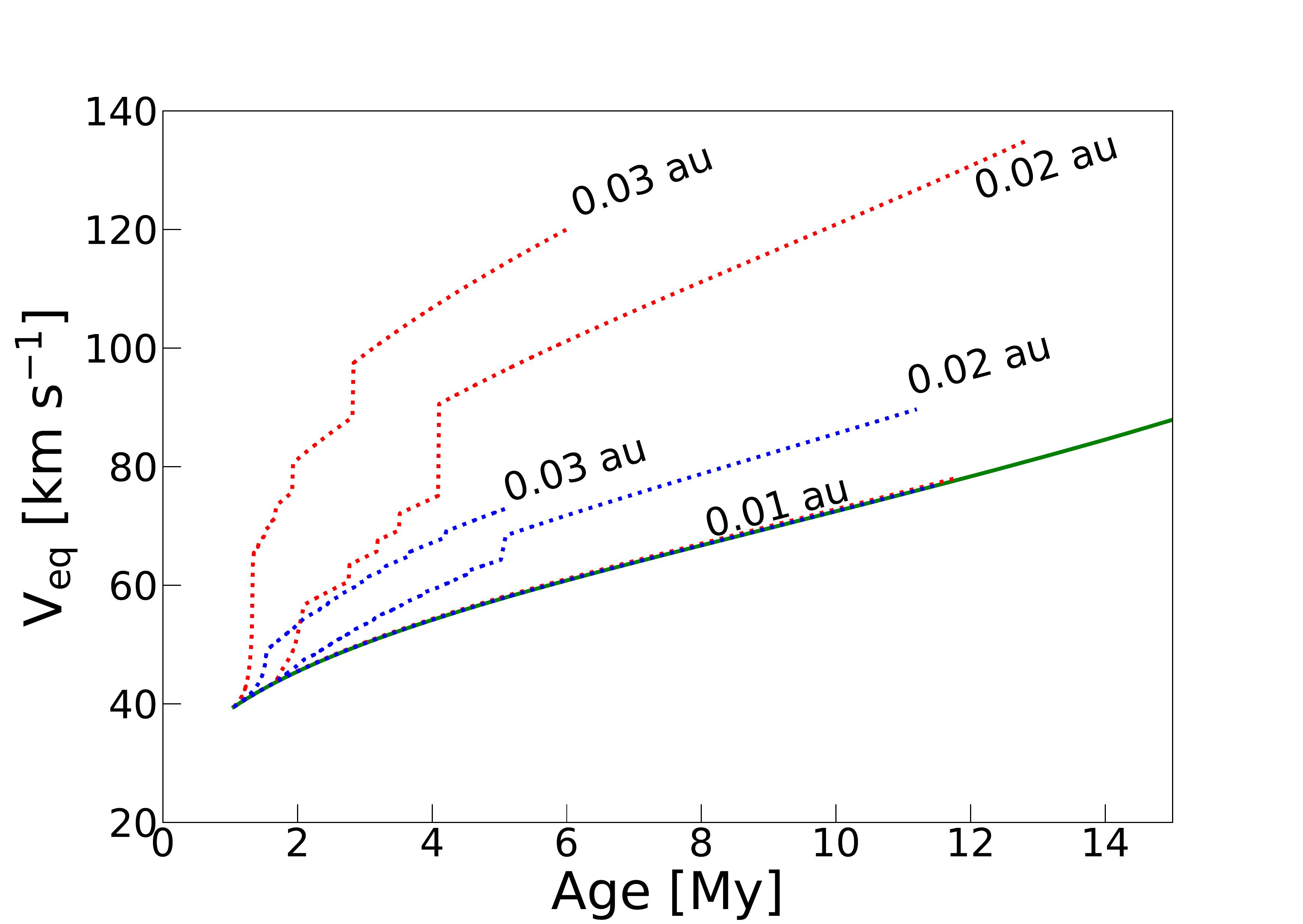}
\caption{Time-evolution of the stellar surface rotation velocity when the orbits of planets with 10 (blue dotted curves) and 30 M$_{ Jup}$ (red dotted curves) shrink, compared to the surface velocity evolution of the star without any planet (continuous green curve). Three dotted curves are shown per planet for three initial distances from the star: The lower curves correspond to 0.01 au, and these follow the same path whatever the mass of the planet, coinciding with the green curve. The middle and upper curves correspond to initial distances equal to respectively 0.02 and 0.03 au.  
}
\label{Veq1030}
\end{figure}

\textit{The intermediate-mass planets ($\sim 0.02$ M$_{ Jup}$ < m$_{ pl} < \sim 2.5$ M$_{ Jup}$):} 
In this domain, both tides and evaporation have significant effects. Interestingly, this is also the mass domain where the evaporation rate goes through a maximum. Indeed, as shown in the right panel of Fig.~\ref{Pote}, the evaporation rate becomes maximum for a mass around 124 M$_{ Earth}$, or 0.39 M$_{ Jup}$, where the mass radius relation (see the left panel of Fig.~\ref{Rpl}) changes slopes.
For the 0.1 M$_{ Jup}$ planet, we see that regardless of whether the orbit shrinks or expands due to tides, as long as the initial distance is below 0.05 au, the planet is completely evaporated on timescales shorter than a few 100 Myr. 
A main difference between the 1 M$_{ Jup}$ planet and 
the 0.1 M$_{ Jup}$ one is the fact that this more massive planet escapes complete evaporation if its initial distance is larger than around 0.034 au. As seen previously, only when the planet orbit shrinks, we have a complete evaporation. The cases where the orbits expand show very little mass loss and very small changes of the radius even after a few Gyr.
The sharp drop in the evaporation rate shown in the panel of the last column in the second row of Fig.~\ref{figOrMa} occurs at the transition between the saturated and unsaturated regime for computing the X-ray flux. This occurs at an age around 300 Myr (see the left panel of Fig.~\ref{figVFR}). The evaporation rate drops to a value of 10$^9$ g s$^{-1}$, which is $ 1.7 \times 10^{-11}$ M$ _{ Jup}~yr^{-1}$. 
The case of the 2 M$_{ Jup}$ planet shows many qualitative similarities with the case of the 1 M$_{ Jup}$ planet. The evaporation rates are smaller since the gravitational potential well is deeper for more massive planets.

\textit{The high-mass planets (m$_{ pl} > \sim 2.5$ M$_{ Jup}$):}
Let us remind that above 2.5 M$_{ Jup}$, the present evaporation laws
cannot be applied, in principle. However, these laws lead to very small evaporation rates, as can be seen looking at the panels of the fourth row in Fig.~\ref{figOrMa}. Hence, the evolution in these cases is similar to 
models without evaporation, and this is the domain where tides dominate. 
Planets with initial distance below around 0.03 au have their orbits strongly affected by dynamical tides and these planets are engulfed by the star. Such an engulfment process produces a significant spin-up of the star. We can see in Fig.~\ref{Veq1030}, that for the 30 M$_{ Jup}$ mass planet starting its evolution at 0.03 au, the surface velocity is increased by a factor of two\footnote{ 
In Fig.~\ref{Veq1030} the evolution of the velocity of the star is not shown after the time of the engulfment.}.


\begin{figure*}
\centering
\includegraphics[width=.5\textwidth, angle=0]{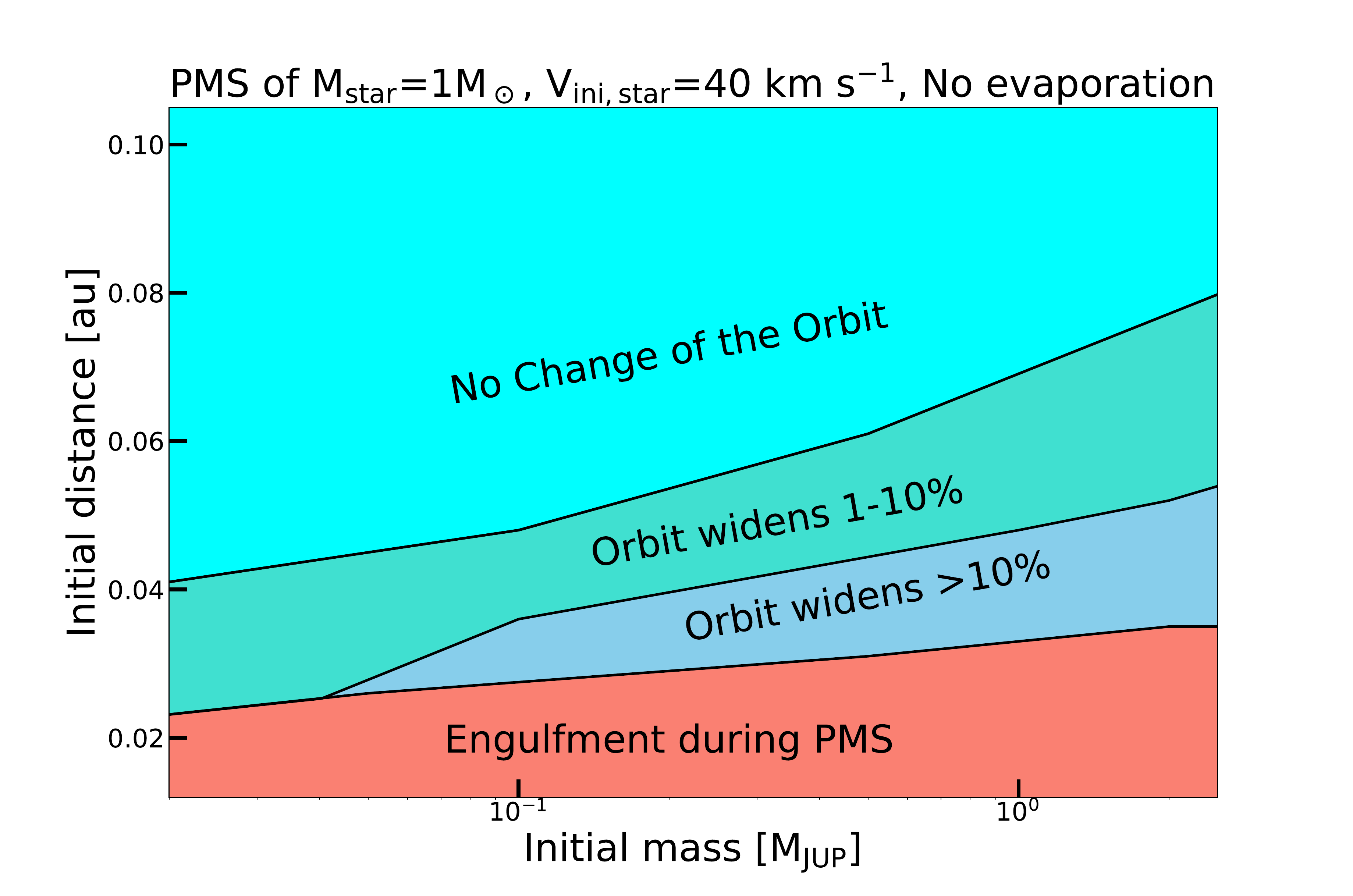}\includegraphics[width=.5\textwidth, angle=0]{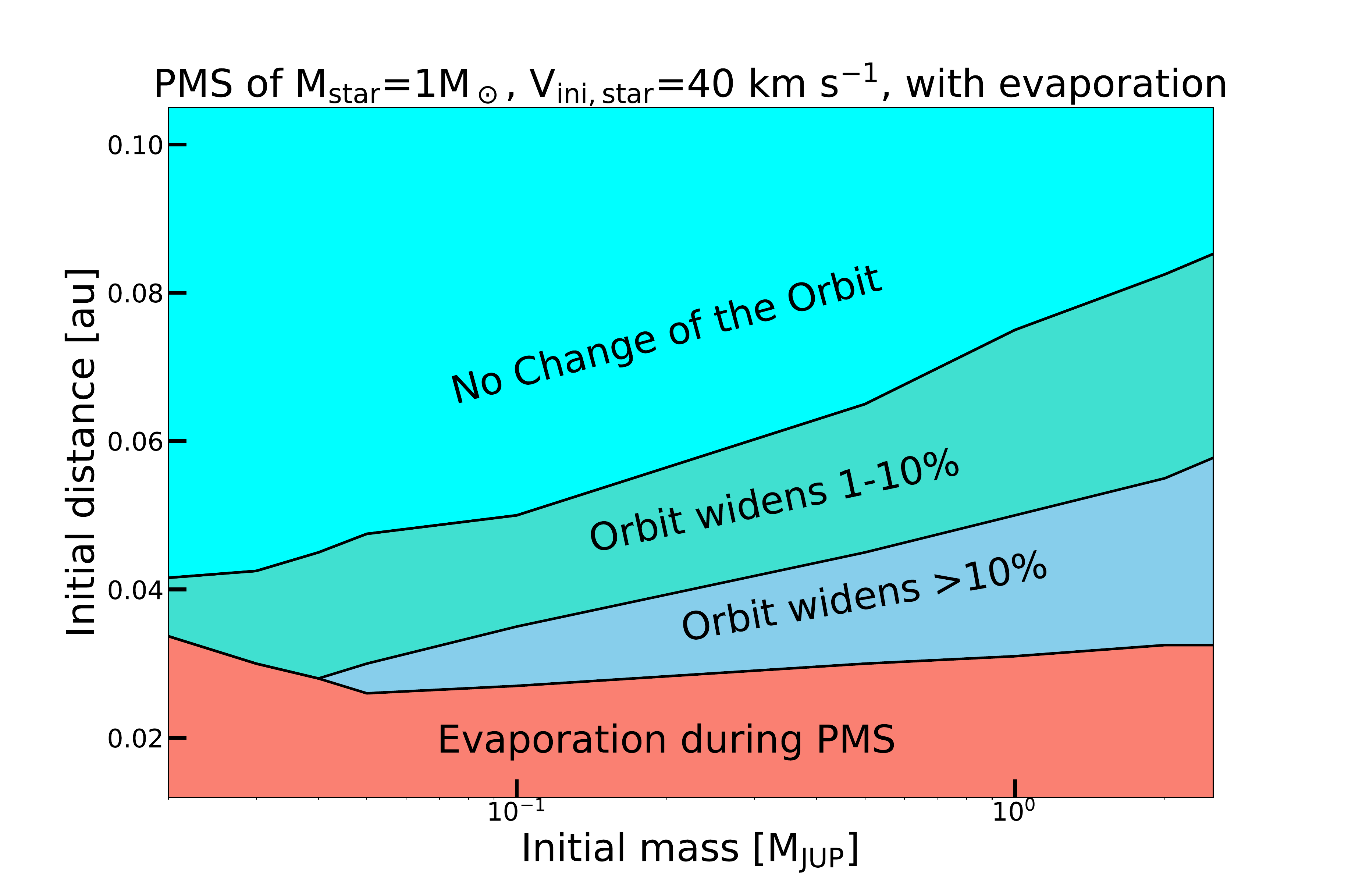}
\caption{Fate of planets of different masses (in Jupiter masses), starting their evolution at various distances (in au) from their host star during the PMS phase.
The left and right panels show respectively, the fate of planets without accounting for the evaporation process and with account of the evaporation process.
Planets in the salmon region are engulfed or evaporated.For low-mass planets, evaporation usually occurs before engulfment, and vice-versa for high-mass planets. Therefore, the transition of the fate happens somewhere in the intermediate-mass region.
Planets in sky-blue regions and in turquoise regions have their orbits widened by tides.
}
\label{compevap}
\end{figure*}

In Fig.~\ref{compevap}, we compare the fate of planets during the PMS phase, with and without evaporation. We see, as expected, that the region that is the most affected by the evaporation process is the low-intermediate planet mass range (planets with a mass below about 0.04M$_{ Jup}$). In this mass domain, a main consequence of the evaporation process is that it enlarges significantly the domain of initial conditions leading to 
the disappearance of the planet (or at least of the planet's atmosphere) as compared to the corresponding domain obtained by considering only tides and no evaporation. Typically, for a planet with an initial mass of 0.02 M$_{ Jup}$, with no evaporation process, the planet avoids engulfment if its initial distance is larger than about 0.025 au. With evaporation,
the planet avoids evaporation if the initial distance is above 0.035 au. In Sect.~\ref{sec:5}, a comparison of the present results (of planets or at least of planets with an atmosphere) is made with those obtained by \citet{Kuro2014}.

\subsection{Evolution beyond the Pre-Main-Sequence phase}

\begin{figure*}
\centering
\includegraphics[width=.5\textwidth, angle=0]{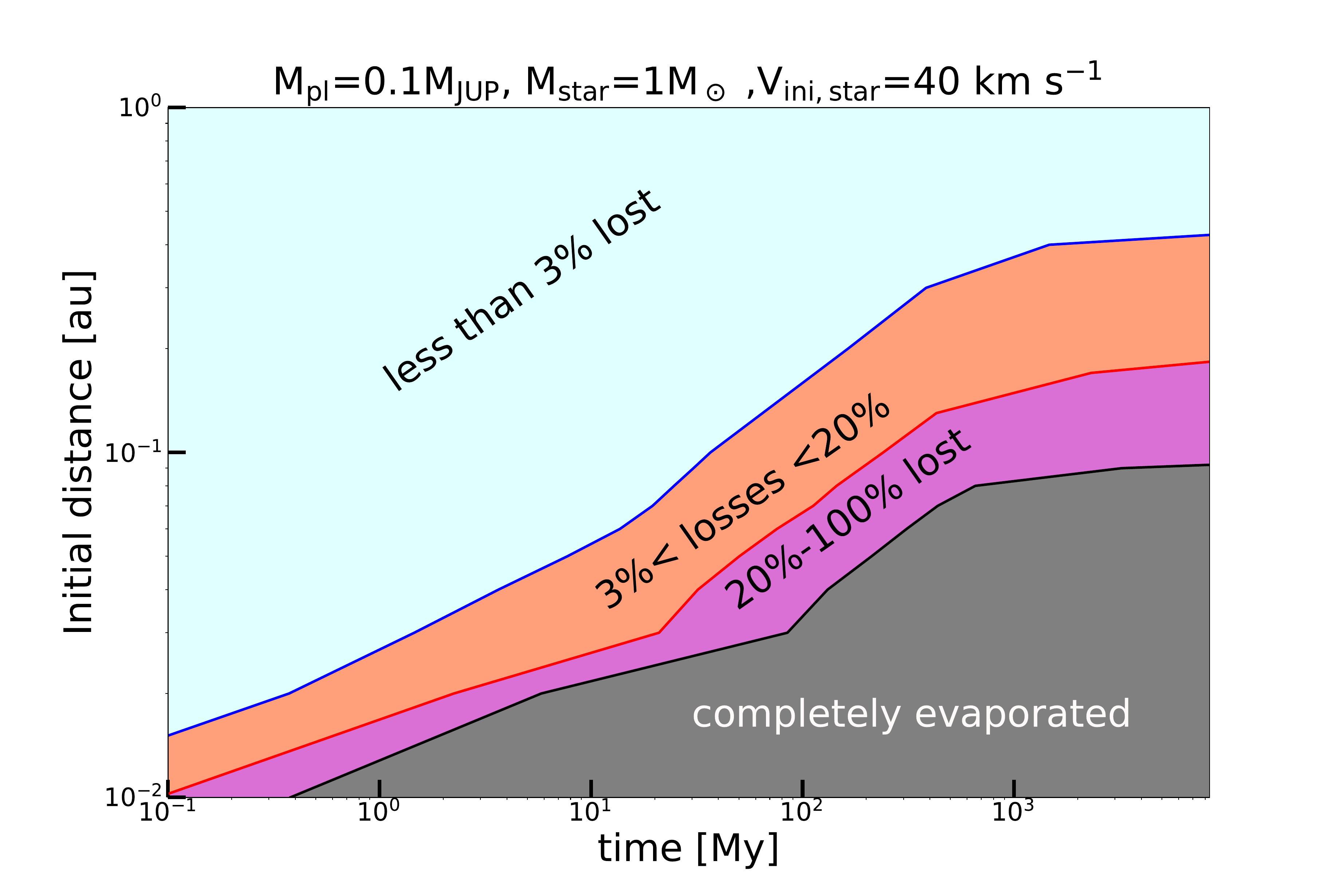}\includegraphics[width=.51\textwidth, angle=0]{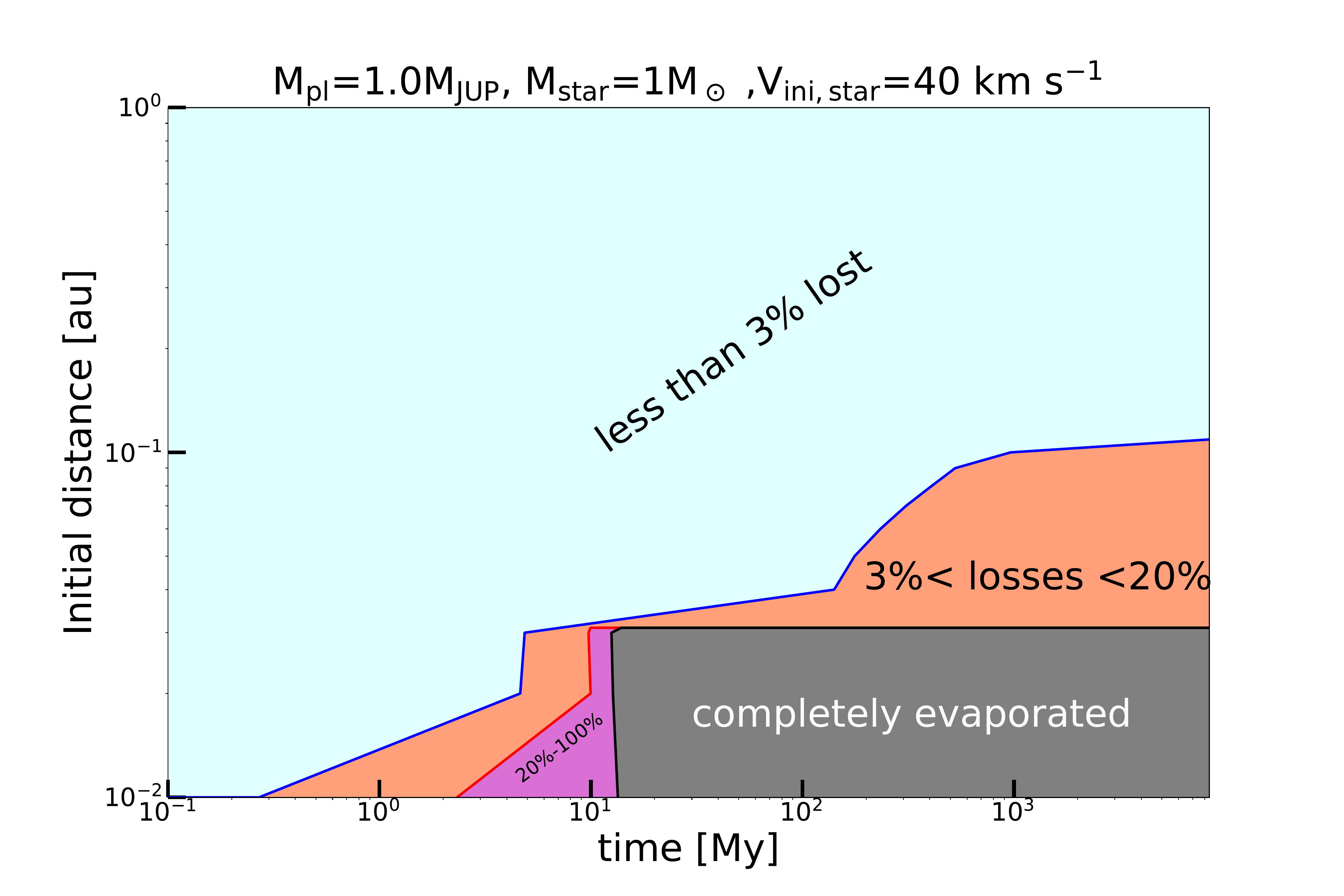}
\caption{Evolution as a function of time of the minimum initial distance between
the planet and its host star above which the planet survives (limit between 
the grey and pink areas), the planet loses less than 20\% of its initial mass 
(limit between the purple and salmon area), the planet loses less than 3\% of its initial mass. The different panels show these limits for different values of the initial mass of the planet. The initial rotation of the star is 40 km $^{-1}$ at 30 Myr before the ZAMS.}
\label{figDisTime}
\end{figure*}

In Fig.~\ref{figDisTime}, we show how the mass of the planets changes as a function of time depending on the initial distance between the planet and the star and for two different initial mass planets. A much larger range of initial distances is considered, compared to Figs.~\ref{figOrMa} and \ref{compevap}, as well as a much larger time domain (the PMS phase ends approximately at an age equal to 40 Myr). The domain of disappearance of the planet (the limit between the grey and the purple area) increases significantly with time for planets with an initial mass equal to 0.1 M$_{ Jup}$ planets. This is, of course, a consequence of the fact that more time allows for more evaporation, which implies that the planet, to avoid becoming completely evaporated by a given time, has to begin its evolution at a greater initial orbital distance. 

We see that for this mass, the limits between the diverse colored areas increases significantly up to around 300 Myr. After this period, as mentioned before, the X-ray flux decreases, and hence the evaporation rate too, causing the limits to vary less rapidly with time than in the period that precedes. 

In the case of the 1 $M_{Jup}$ mass planet, all the cases beginning their evolution at a distance below 0.03 au are completely evaporated at more or less the same age (see the second row, second column panel in Fig.~\ref{figOrMa}). Interestingly, for this specific planet mass, complete evaporation occurs only during the PMS phase, otherwise it never occurs. This is reflected by the fact that the upper limit of the grey zone is a nearly horizontal line. If a planet begins its evolution at a distance a bit larger than about 0.03 au, then it will escape complete destruction during the whole MS phase. We see that starting at an orbital distance above 0.03 au will make the time during which tides can shrink the orbit very brief. Indeed, at this distance, the orbit rapidly crosses the corotation radius. After this crossing, the orbital distance expands, thus saving the planet from the loss of its atmosphere as it spirals away from the star.

Comparing the two panels of Fig.~\ref{figDisTime} shows that increasing the initial mass of the planet reduces the zones where the planet mass changes. This simply reflects the fact that a more massive planet, in order to be evaporated at a given age, must start its evolution nearer to its host star. 
This effect is particularly significant when going from 0.1 to 1 M$_{ Jup}$ mass, {\it i.e.} through the mass range where the mass-radius relation for the planet changes slope. Beyond a mass of around 0.4 M$_{ Jup}$, the gravitational potential increases more rapidly with the mass than in the lower mass domain. This makes the planets more resistant to an evaporation process and thus, significantly restricts the conditions for losing a given amount of mass.

\begin{figure}[!h]
\centering
\includegraphics[width=.5\textwidth, angle=0]{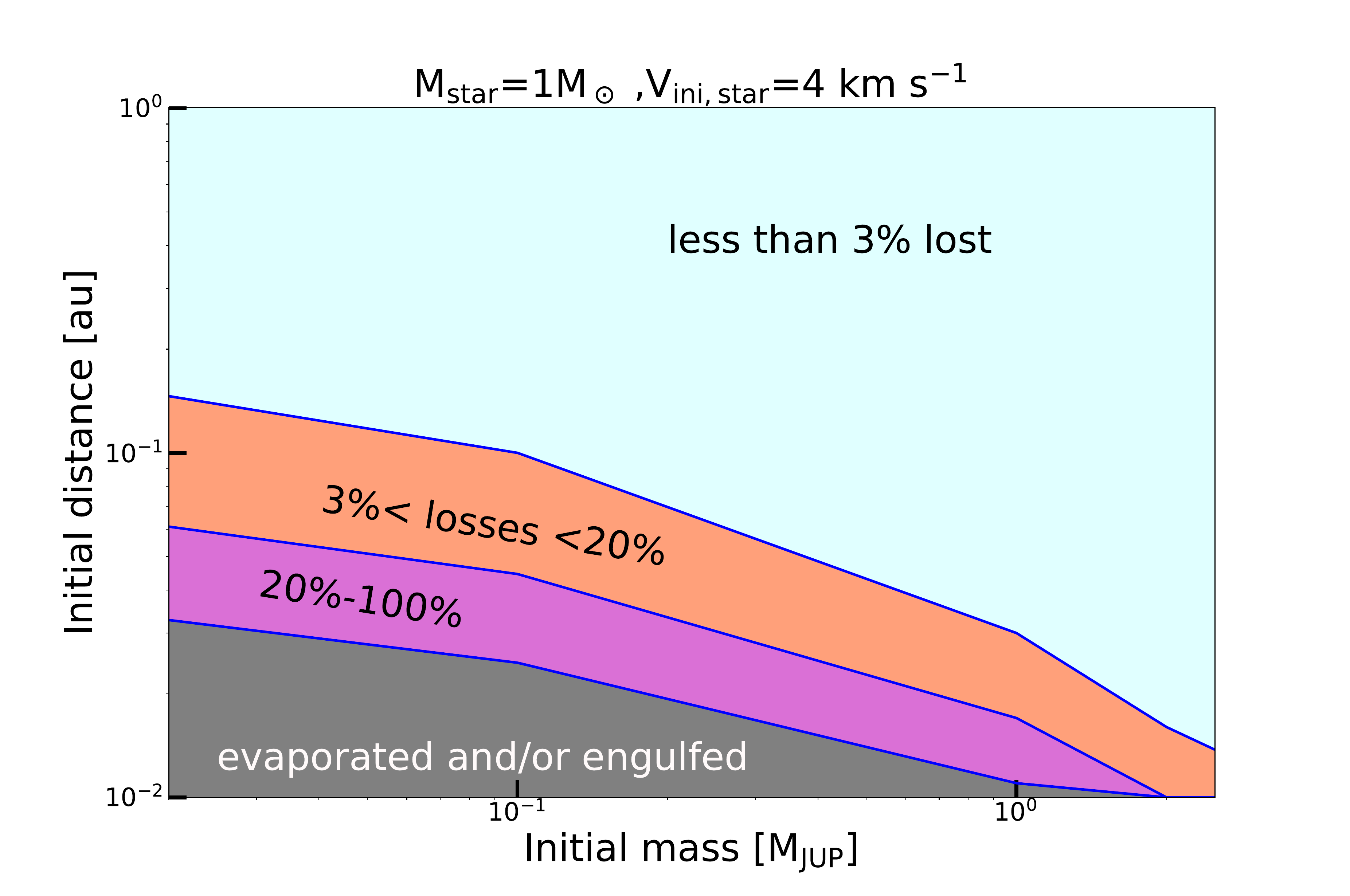}
\includegraphics[width=.5\textwidth, angle=0]{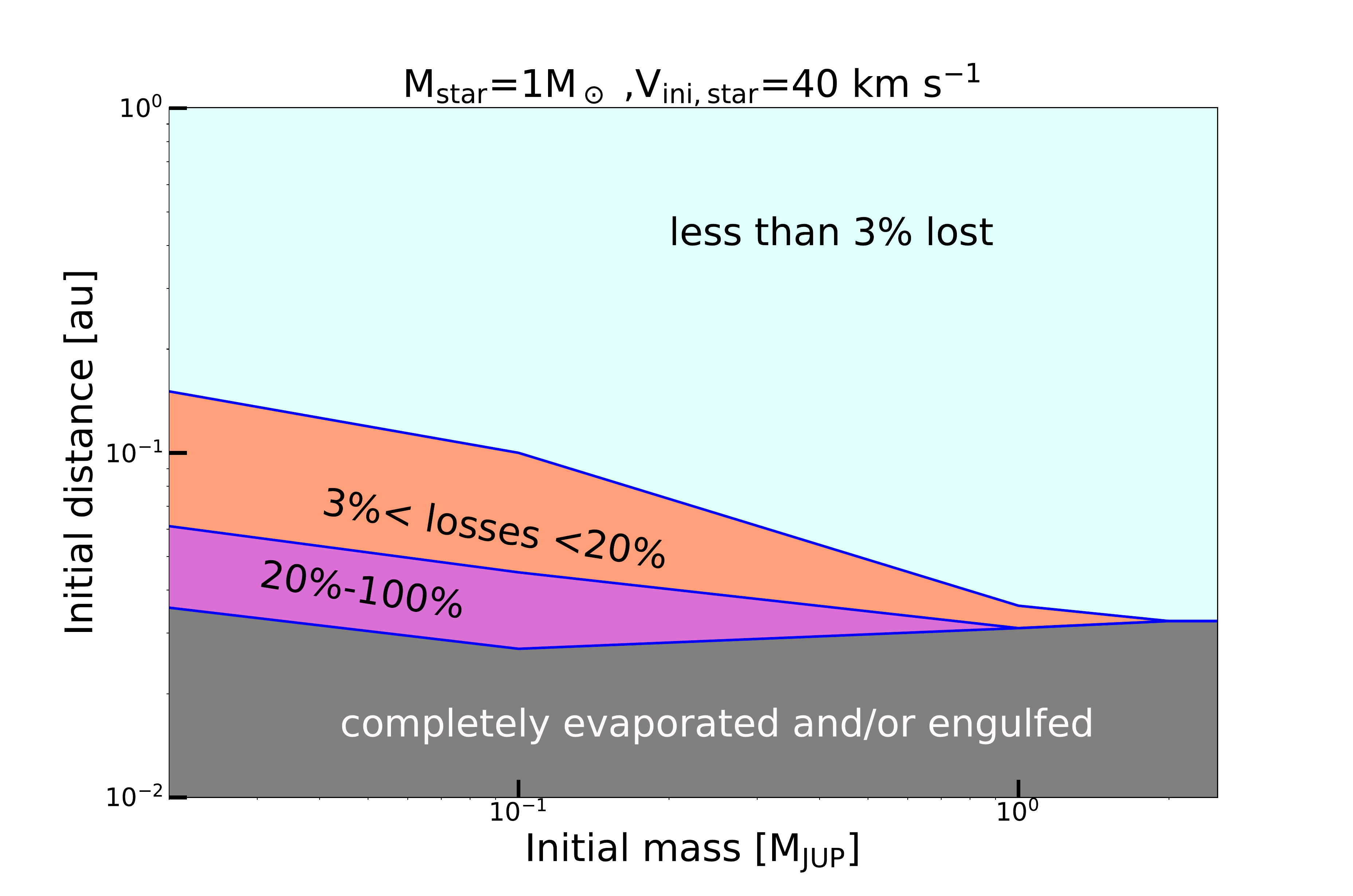}
\includegraphics[width=.5\textwidth, angle=0]{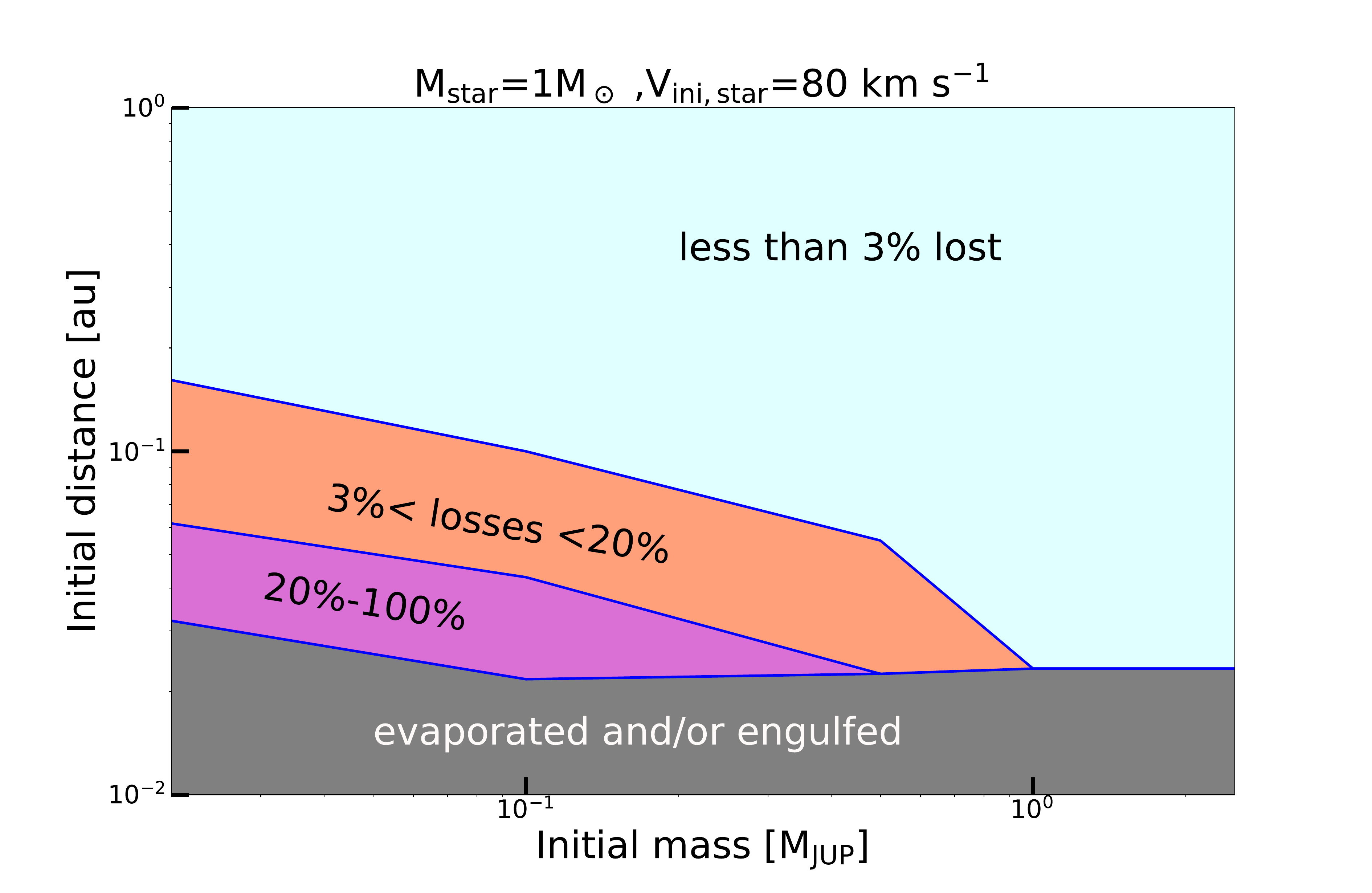}
\caption{Evolution as a function of the initial mass of the planet of the minimum initial distance between
the planet and its host star above which the planet survive the Pre-Main-Sequence phase (limit between 
the grey and pink areas), the planet loses less than 20\% of its initial mass 
(limit between the purple and salmon area), the planet loses less than 3\% of its initial mass (limit between the salmon and light blue areas). {\it Upper panel:} The initial rotation of the star, taken at an age equal to 30 Myr before the ZAMS is 4 km s$^{-1}$. {\it Middle panel: } Same as upper panel, but for an initial rotation of the star equal to 40 km s$^{-1}$.  {\it Lower panel: } Same as upper panel, but for an initial rotation of the star equal to 80 km s$^{-1}$.}
\label{figureVC}
\end{figure}

\subsection{Impact of the initial stellar rotation}

In Fig.~\ref{figureVC}, we compare for three different initial rotations of the star, how the minimum distance for an evaporation and/or an engulfment during the PMS phase varies as a function of the planet mass.

We note that for planet masses below about 0.1 M$_{ Jup}$, changing rotation has little effects. This is in line with two facts: First, for this planet range, the evaporation process dominates. Second, the XUV flux is saturated (see the left panel of Fig.~\ref{figVFR}), and thus does not depend on rotation.

In the high-mass planet range (for masses above about 0.1 M$_{ Jup}$),
the initial stellar rotation has a significant impact. As a numerical example, a 1 M$_{ Jup}$ planet orbiting a 1 M$_\odot$ star with an initial rotation of 4 km s$^{-1}$, 30 Myr before reaching the ZAMS, can survive if its initial distance is larger than about 0.01 au.
When the initial stellar rotation is 40 km s$^{-1}$, the planet should have an initial distance larger than 0.03 au. 
This is a consequence of the dependence of the dynamical tides with the rotation rate of the star.  

If we compare the situation where the planet orbits a moderately fast  rotating star (40 km s$^{-1}$) to the one in which it orbits a fast rotating model (80 km s$^{-1}$, compare the middle and lower panels of Fig.~\ref{figureVC}), 
we observe for the faster rotating star,
a decrease of the minimum distance for engulfment in the planet mass range above 0.1 M$_{ Jup}$ (engulfment due to tides dominates over evaporation in the high-mass planets range). This is at first sight, surprising, because faster rotation implies stronger dynamical tides and thus, more efficient shrinking of the orbit. But at the same time, faster rotation also implies that the corotation radius is pushed inward, reducing thus the zone where dynamical tides shrink the orbit and favour an engulfment.

\section{The different types of evolution}\label{sec:4}

Very generally, close-in planets 
show different types of evolution depending on the relative magnitudes of three timescales. They are
the evaporation timescale, $\tau_{ evap}=M_{ pl}/\dot M_{ pl}$, the tidal timescale, $\tau_{ tides}=(a/\dot a)_{ti}$ with $a$ being the orbital distance, and the stellar evolutionary timescale $\tau_{ star}$, which is more rapid during the PMS phase (mainly the Kelvin-Helmholtz timescale) than during the MS phase (the nuclear timescale, although as explained below for the specific question addressed here, the magnetic braking timescale is certainly more relevant). We briefly summarise the different types of evolution
with emphasis on how stellar rotation and feedback affect them, since these effects
are the focus of the present study. The different cases described below apply to different initial planet mass - orbital distance domains. We order their description beginning with the domain containing the most massive and distant planets, and finishing with the least massive and closest ones.

\subsection*{Constant orbit and mass evolution ($\tau_{ evap}$ \& $\tau_{ tides} >> \tau_{ star}$)}

This evolution occurs for distant and high mass planets, so that neither the orbit nor the mass of the planet are significantly affected by tides and evaporation. The evolution of the planet in such cases is not affected by the rotation of the star either, although however, the minimum mass (given an initial distance) or the minimum distance (given an initial planet mass) for such evolution to occur is shifted to larger values when the stellar rotation increases.

\subsection*{Constant mass evolution ($\tau_{ evap} >> \tau_{ star} > \tau_{ tides}$)}\label{cm}  

This occurs for sufficiently large mass planets
having an initial orbital distance below the corotation radius, so that tides shrink the orbit. The upper limit of the initial orbital distance for this type of evolution to occur is a bit below the initial corotation radius, where the orbit plunges just fast enough such that it always remains below the corotation radius, which itself also shrinks with time during the PMS phase. Thus, the orbit keeps plunging till it reaches the minimum orbital distance above which dynamical tides become active, $a_{ min}$. In other cases, if the planet begins its evolution below $a_{ min}$, then initially for a while,
the orbit is not changed. At a given point, however, the orbital distance will become equal to $a_{ min}$, because the star contracts during the PMS phase, so its rotation increases and $a_{ min} \propto \Omega_{ star}^2$ decreases. From this stage on, in all cases, the orbital distance 
will remain in the vicinity of this limit (see, as general examples of this effect, the orbits beginning at 0.02, 0.03 au in the first panels of the second, third and fourth rows, in Fig.~\ref{figOrMa} \footnote{In the results shown in this work, the orbital distance and $a_{ min}$ curves are not exactly coinciding only due to time discretization in the numerical procedure, otherwise they would.}). This happens because, every time the orbital distance drops a bit below $a_{ min}$ due to the tides, the dynamical tides switch off and the orbital distance momentarily stops changing, until
the $a_{ min}$ limit reduces and crosses the orbit, switching the dynamical tides back on, and the process thus continues. 

Engulfment will occur for these planets during the PMS phase. In those cases, tides may spin-up the star significantly and feedback may be important enough to be accounted for. The domain for this type of evolution depends on rotation in two ways: First, through the strength of the dynamical tides that increases when rotation increases and second, through the determination of the upper (corotation radius) and lower distances ($a_{ min}$) where dynamical tides
can shrink the orbit. Both these distances are shifted to closer-in distances when rotation increases. In order to have the evolution described here, one needs to have this region sufficiently close to the star so that the tidal timescale is short enough. 
In general, increasing the initial rotation will
increase the domain in the plane of initial mass vs. initial distance where this type of evolution occurs. However, beyond some limiting rotation, we expect to have a reduction of this domain due to the fact that
tides will anyway be high enough to produce an engulfment during the PMS phase, but the domain where the tides are active becomes smaller (see Fig.~\ref{figureVC}). Here, feedback effects can also be important since, during the orbit shrinking, the star will spin-up and thus, the strength of tides increases, while the domain where tides are active shrinks and shifts closer to the star.

\subsection*{Orbit and mass evolution ($\tau_{ evap}$ \& $\tau_{ tides} < \tau_{ star}$)}\label{om}

This evolution characterizes the domain of sub Neptune and hot Jupiter planets. In these cases, evaporation can occur either during the PMS phase or during the early MS phase when the star still has a high enough rotation to emit a saturated XUV luminosity. This is an important domain because it impacts a large part of the survival zone shown in Fig.~\ref{EXO}.

If the planet begins its evolution at an initial orbital distance which is above the upper limit mentioned in the previous subsection, then the orbital distance will eventually increase due to tides after the corotation radius shrinks below it. Therefore, the star will spin-down, which will not impact the evaporation efficiency during the PMS phase since the star contracts rapidly during this time, which keeps its XUV photon emission saturated even if some angular momentum is lost at the profit of the expanding planetary orbit. The irradiation flux will, however, decrease because the orbital distance increases, which will therefore also increase the evaporation timescale.
If not too massive and not too distant from their host stars, these planets
can still lose significant amounts of their initial mass over periods of a few 100 Myr. Examples of such an evolution in the present simulations are the cases of the 0.1 M$_{ Jup}$ planets with initial orbital distances of 0.03 and 0.04 au, which are evaporated around 100 Myr (calculated for an extrapolation of the cases shown in the first row of Fig.~\ref{figOrMa}). For higher values of the initial rotation of the star, the PMS phase described above will not be affected much. However, the period during which the star emits high energy photons after the PMS phase will increase. This decreases the evaporation timescale. Not only is the initial rotation of the star important, but also the efficiency of the internal angular momentum transport mechanism. A less efficient transport mechanism than the one considered here may reduce by a factor of three the duration of the XUV radiation emissions.

If instead, the planet begins its evolution below the orbital distance upper limit mentioned in the previous subsection, then the orbit shrinks. Shrinking the orbit causes the planet to experience a higher irradiation flux, and thus,
both timescales, $\tau_{ tides}$ and $\tau_{ evap}$, decrease. In our computations, we obtain that the evaporation effect dominates over tides in these cases, obliterating the planet before engulfment. This is seen in, for instance, the cases of the 1 M$_{ Jup}$ mass planet with initial orbital distances of 0.02 au and 0.03 au, in the third row of Fig.~\ref{figOrMa}. 

Changing the initial rotation has modest effects on this type of evolution for the reasons already mentioned above.
Feedback effects of the tides on the stellar rotation are modest because of the fact that this type of evolution involves planet masses
and distances that provide modest amounts of angular momentum, and only part of it is transmitted to the star through tides because of the evaporation process.

After the early Main-Sequence phase that lasts a few 100 Myr for a star like the Sun, the planets that have survived no longer evolve much, at least for those whose mass is below the critical mass for activating tidal dissipation in radiative zones. It is only when the star becomes a red giant that new important changes, mainly due to equilibrium tides, can occur. Thus, the surviving planet here will follow a constant orbit and mass evolution during the MS phase.

\subsection*{Mass evolution ($\tau_{ evap} << \tau_{ tides} < \tau_{ star}$)}

This evolution occurs for planets that are the most close-in and have the lowest mass in the ranges considered in our work. The planet loses its mass before tides can modify its orbit. For instance, in the present models, we obtain that a 0.1 M$_{ Jup}$ planet with an initial distance of 0.01 au would have such an evolution (see first row of Fig.~\ref{figOrMa}). Stellar rotation has little impact on this type of evolution. Even starting from a very small initial rotation, since the star is contracting during the PMS phase, it will rapidly spin-up and emit high energy photons at the saturation limit. Since tides are not important, rotation will also have no impact on their action.

\section{Comparisons with observations}\label{sec:5}

\subsection{Neptunian desert}

\begin{figure}[!h]
\centering
\includegraphics[width=.49\textwidth, angle=0]{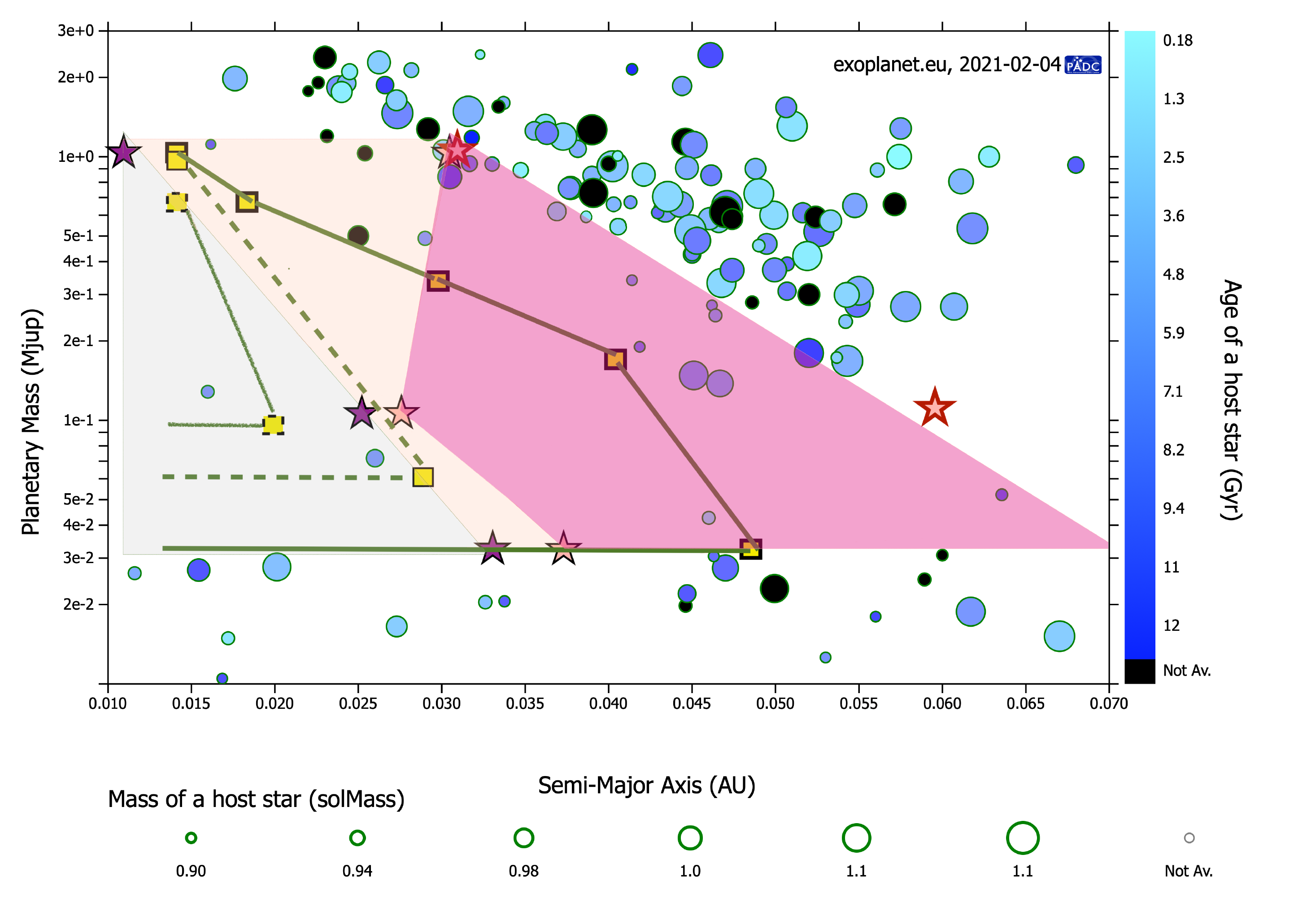}
\caption{The data points are masses ($M\sin i$) and semi-major axis for confirmed planets orbiting stars with masses between 0.9 and 1.1 M$_\odot$ (diagram made using the database exoplanet.eu). The light grey shaded area indicates the Neptunian desert zone after 40 Myr as predicted by our models for our slow rotating case (V$_{ ini}$= 4 km s$^{-1}$). The light pink zone shows the enlargement of that Neptunian desert region just after 40 Myr as predicted by our models for our moderate rotating case (V$_{ ini}$= 40 km s$^{-1}$). The magenta shaded area shows the enlargement of that zone after
300 Myr. 
For comparison, also plotted are the lines corresponding to the minimum survival masses of planets with core masses of 0.03 M$_{ Jup}$ (solid line), 0.06 M$_{ Jup}$ (dashed line) and 0.09 M$_{ Jup}$ (fuzzy line) at an age of 10 Gyr obtained by \citet{Kuro2014}. The corresponding core masses are shown by horizontal lines.}
\label{EXO}
\end{figure}

In Fig.~\ref{EXO}, the masses ($M\sin i$) and distances of confirmed exoplanets orbiting solar mass stars are shown. We have overplotted shaded areas indicating the regions that, according to our simulations, should be devoid of planets either because of evaporation or engulfment. We have focused here on the planet mass domain between 0.03 and 1 M$_{ Jup}$
(i.e. between 10 and 310 Earth masses), which lies well inside the domain where the evaporation law adopted here has been applied in previous works \citet{Salz2016}.

The grey and salmon shaded areas show the Neptunian desert after 40 Myr when the initial stellar rotation is 4 and 40 km s$^{-1}$ respectively. 
Increasing the initial stellar rotation enlarges the desert zone, especially for the most massive planets of the mass domain considered here. This is mainly an effect
linked to the fact that when the rotation of the star increases, tides are stronger (given a fixed initial distance). For those planets below the 
corotation radius, this will favor planet engulfment. For the lower mass end considered here, increasing rotation also enhances the tides which, for the orbit shrinking planets, favors evaporation.

When systems older than about 40 Myr are considered (here up to 300 Myr and and for an initial rotation of 40 km s$^{-1}$), the desert domain is enlarged, mainly because it gives more time for the planet to be irradiated by the high energy photons of the star. Changing rotation has an impact on this region, since it modifies the duration of the period during which the XUV flux is saturated.

A very interesting feature to note is that the enlargement of the desert zone when the age increases
is much larger for the 0.1 M$_{ Jup}$ planet than for the 
1 M$_{ Jup}$ one. This is because, for a low-mass planet, the evaporation timescale at a given distance is much shorter than for a high-mass one. This effect imposes thus that the right frontier of the magenta zone goes towards higher initial distances when the mass of the planet decreases.
We see that overall, the zone devoid of planets predicted by the present simulations is reasonable with respect to the observed distribution of planets in this plane. Of course, much more refined comparisons are needed to incorporate more realistic planet models. 

For comparison, the minimum survival masses of planets with different core masses observed at an age of 10 Gyr, as predicted by \citet{Kuro2014}  are shown. These authors considered more sophisticated interior planet models, but they did not consider any effects of tides. This is mainly the reason why their survival limit is at very small distances for the their 1 M$_{ Jup}$ model. They adopted the XUV evolution models given in \citet{Ribas2005}, based on observations of solar analog stars, thus not accounting for the impact of different initial rotations and for exchanges of angular momentum between the planet orbit and the stellar rotation.
We see that our results significantly enlarge the zones devoid of planets. On one hand this comes from the fact of accounting here the impacts of tides. On the other hand we may here
somewhat overestimate the desert zone because we do not allow for the existence of rocky cores.

We note that a more complete interpretation of the observed Neptunian desert also requires accounting for high-eccentricity orbital migration (\cite{Owen2019}). Although the present simulations do not account for eccentric planet orbits, we show here that simply accounting for tides, evaporation and stellar evolution is still a good approximation to explain the observed results. 

\subsection{Orbital period upper limit for bare-core planets}

\begin{figure}[h]
\centering
\includegraphics[width=.49\textwidth, angle=0]{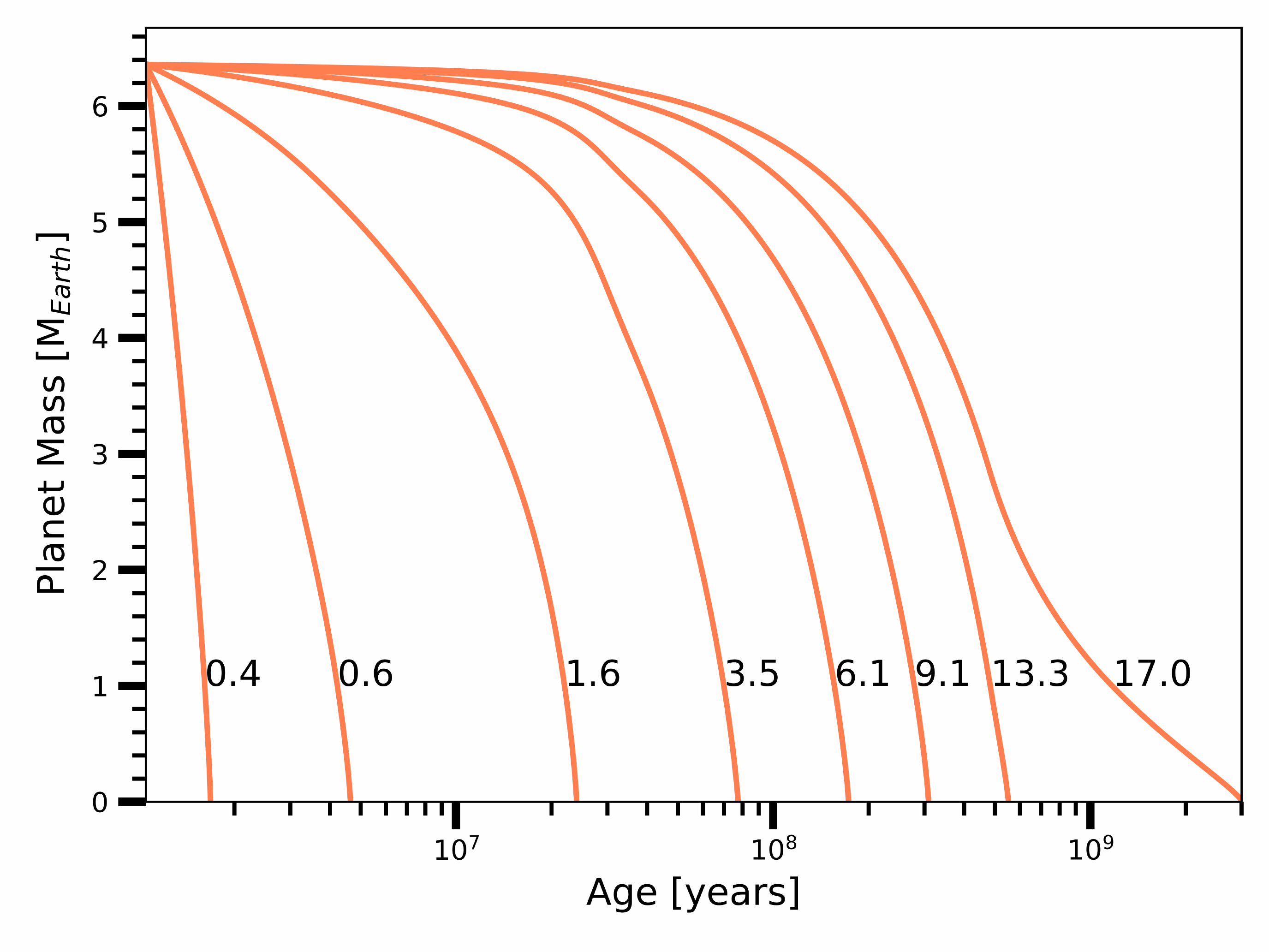}
\caption{Evolution of the planet masses as a function of the age. The initial mass of the planets is 0.02 M$_{ Jup}$, being the smallest in the mass range considered in our study. The stellar model is for a 1M$_\odot$ star beginning its evolution with a surface velocity of 40 km s$^{-1}$, 30 Myr before the ZAMS. Each curve is labeled by the orbital  period of the planet at the end of the evolution (see text), in units of Earth days. From the shortest to the longest orbital period, the initial distance has the following respective values in au: 0.0100, 0.0150, 0.0250, 0.0450, 0.0650, 0.0850, 0.1100, 0.1294.
}
\label{RPeriod}
\end{figure}

From the right panel of Fig.~\ref{compevap}, we can deduce the trend that planets with the smallest initial mass in the range we consider, can begin their evolution at the farthest initial orbital distance from the star, such that at the end of the evolution i.e. at a given final age of the system, the planets just lose their entire atmosphere due to evaporation, while still being as far away from the star as possible. In other words, for star-planet systems of a given final age, computations of the final orbital period of the planets beginning their evolution with the smallest initial mass considered in our study, which just lose their atmosphere due to evaporation at that age, would be an upper limit on the orbital period as predicted by the present simulations, above which small-radii/bare-core planets cannot be observed. 

In Fig.~\ref{RPeriod}, we make this computation, and find that the upper limit of the orbital period for observing bare-core planets around solar-type stars of a few Gyr age, is around 17 (Earth) days. This number is in surprisingly good agreement with the locus of the small-radii/bare-core population constituting in part the "radius valley" feature i.e. the bimodal distribution of exoplanets observed by the Kepler mission, in a scatter plot of planet radius vs. orbital period (see \cite{vaneylen2018}). We find that both in observations, as well as in planet population simulations accounting for more detailed planet physics than we do (\cite{Owen2017}), the bulk of the small-radii/bare-core population lies just below the upper limit predicted by our code, indicating this limit to mark, as one would expect in principle, a boundary of the radius valley feature. 

However, since we do not model the interior structure of planets in the present work, we cannot say anything about the radii of the bare-core planets left over after complete evaporation of their atmospheres. Hence, we can only find numerical limits/boundaries along the orbital period (horizontal) axis, but not along the planet radius (vertical) axis of the planet radius vs. orbital period scatter plot.

The small fraction of observed bare-core planets that do lie above our predicted orbital period upper limit, must have followed an evolution or initial conditions different from the ones covered by the present simulations, implying, without much surprise, that more physics is involved in star-planet systems than is considered here. On the other hand, the very good agreements, taken at face-value, imply that many star-planet systems in the observations may have followed an evolution similar to the ones predicted by our simple, yet comprehensive models.

\section{Advantages and limitations of the present approach}\label{sec:6}

The main advantages of the present consistent treatment are the possibilities to study the impact of different initial rotations on the evolution of the planet mass and orbit, of accounting for the feedback of these changes on the stellar rotation, and thus on the tides and evaporation rates.

A main present limitation is
the rather schematic interior model used for the planets. 
In this work, we use a mass-radius relation that fits direct observational exoplanet data and numerical simulations of planet formation \citep{bashi2017}, therefore being a valid first approximation. What is shown here is that, at least in some planet mass domains, the impact of changing the initial rotation of the star can have effects as large, or even larger than considering different planet interior models (see Fig.~\ref{EXO}).

We did not consider multiple planets around the star and therefore, any interactions between them, nor the impacts of possible magnetic interactions between the star and planet. Neither did we account for eccentric and/or inclined planet orbits, assuming always a quasi-circular orbit of the planet in the rotation plane of the star.

The limits of the survival zone depend on when the planet arrived at the point defined here as the initial orbital distance to the star. For recall, we consider here a starting point that corresponds to a time of about 40 Myr before the ZAMS. Would we have considered that the protoplanetary disk dissipates at later time, then it would impact the strength of the tides, making them weaker. Indeed, the extent of the convective envelope and the radius of the star decrease as a function of time during the PMS phase. 
The planet may also become a close-in planet while beginning its evolution at much larger distances due to migration and/or dynamical interactions.
A fraction of close-in planets could indeed have migrated long after their formation via Kozai migration, thus delaying their atmospheric evolution and evaporation by several Gyr (\cite{Attia_2021}).

Some uncertainties pertain the XUV stellar luminosity. Recently, \citet{King2021} have argued that the decline in the EUV emission is much slower than for the X-rays and thus might dominate the irradiation of the planet atmosphere after the saturation phase.

We neglected the mass loss by the star and the possible drag forces acting on the planet orbit. These aspects do not appear to be very relevant though. In Fig.~\ref{figlinkSI}, the models from \citet{rao2018} have been computed accounting for these drag forces, and as we can see, they compare extremely well with the present models that neglected these forces.

In a previous section, we briefly mentioned the fact that not only the initial rotation, but also the efficiency of the internal angular momentum transport inside the star has an impact. 
This is certainly a point worthwhile to be studied in a future work.

We focused here on solar mass star models at solar metallicity. Different types of stars have different radiative and convective regions, which change the way tides act. They also exhibit different XUV radiation emission, modifying the photo-evaporation process. Also, the evolution of their rotation can be different depending on the importance of the outer convective zone.
Exploration of these effects will be done in a future work.

We consider here two types of tides, viz. the equilibrium and dynamical tides in convective regions. The expressions for these tides are still subject to discussion \citep[see e.g. the discussion in][]{Barker2020}. Also, dissipation of tides in radiative zones due to internal gravity waves has not been accounted for in the present work. However, according to the recent work
cited above \citep{Barker2020}, this mechanism is efficient only for planet mass above a certain critical limit. For planets orbiting 
a 1 M$_\odot$ star, this critical limit is larger than
10 M$_{ Jup}$ for ages below about 3.5 Gyr
\citep[see Fig. 8 in][]{Barker2020}. The results presented above concern lower mass planets in an age domain below the one indicated above. Thus, they should not be affected by the neglect of the tides in radiative zones. In Fig.~\ref{figDisTime},
we show an evolution over a longer period, until an age of 8.4 Gyr, which corresponds to the end of the core H-burning phase for a 1 M$_\odot$ star. The minimum planet mass at that age is 0.4 M$_{ Jup}$ according to \citep{Barker2020}, hence being above 0.1 M$_{ Jup}$.


There are also the uncertainties pertaining to the evaporation rate. Different choices for the heating efficiency $\epsilon$, the factor $\beta$
describing the inflation of the planet outer layers, the distance to the Roche radius etc. will change the results. Here, we have adopted the given set of values because we wanted to study the impacts of both tides and the evaporation process, as well as the impact of the initial rotation of the star. 
Of course, different evaporation laws may bring significant changes to the results.

\section{Conclusions}\label{sec:7}

As mentioned in the introduction, very few studies follow in a consistent way the change of the planet orbit with that of the stellar rotation, despite the fact that these quantities are tightly connected and impact on each other as well as on the planet irradiation and thus, its evaporation rate.

In the present work, using complete stellar models for describing the evolution of the structure, assuming solid body rotation for the star \citep[which appears as a very reasonable assumption, see e.g][]{Garcia2007, Benomar2015,Eggenberger2019}, we explore the consequences of these two effects in a 4D-space domain whose axes are the planet mass (in the range between 0.02 up to 2.5 M$_{ Jup}$ mass), the initial orbital distance (from 0.01 to 1 au), the stellar rotation (from 4 to 80 km $^{-1}$, 40 Myr before the ZAMS), and the age of the system (from the PMS until the end of the early MS phase).

We have checked the continuity of the present results with those obtained using a previous version of our code where the evaporation process was not accounted for.
Despite the simple models we use here for the planet interiors, comparisons with observations shows that at least for sub-Neptune planets, the results obtained here are reasonable. This gives confidence that our study of the impact of the initial rotation of the star and of the interactions between tides, evaporation and stellar rotations have some validity.

We distinguish two effects of rotation: those linked to different values of the initial rotation of the star, and those arising from the change of the stellar rotation due star-planet interactions.

Changing the initial stellar rotation has the greatest effects on two families of evolution:
the Constant-mass-evolution, and the Orbit-and-mass-evolution (Sect.~\ref{om}). In both cases, in general, increasing the initial rotation tends to increase the domain in initial planet masses and orbital distances where this type of evolution occurs. As mentioned earlier, however, beyond some rotation limit, the domain may decrease due to the dependence on rotation of the limits where dynamical tides are active.
The effects of changing the initial rotation are large on the populations of planets expected just after the PMS phase or after the early MS phase. At larger ages,
the efficient wind magnetic braking makes the system forget the initial rotation.

Interaction between tides and stellar rotation are expected to occur in the case
of Constant-mass-evolution and to a lesser extent in the case of Orbit-and-mass-evolution families. This may produce some significant increase of the surface velocity of the star, as has also been shown by \citet{Gallet2018}. The spin-up of the star due to an engulfment may have a significant impact on other, less massive planets also present in the system that, after the engulfment of a massive planet, may feel stronger tides. This is a point interesting to be studied in the frame of the evolution of planetary systems.

We have seen that the spin-up of the star due to tides has little effect in terms of increase of the XUV luminosity since, in the present simulations, when the irradiation is efficient, it is saturated. The effect of tides (when shrinking the orbit) implies an increase of the irradiation because the star-planet distance becomes smaller, and thus, shortens the planet lifetime.

Although we did not study this effect here, we expect that a change in the efficiency of the angular momentum transport inside the star may significantly change the results for those planets whose evaporation occurs during the early MS phase. Adopting a less efficient angular momentum transport than the one in this study may shorten the duration of the phase during which the star emits a high XUV radiation luminosity.

A general conclusion of the present work is that the huge efforts that have been done so far to better describe the planet physics, evaporation process etc. should also be done in the frame of models where the physics of the star is better described.

\begin{acknowledgements} 
We acknowledge the anonymous referee's critical comments which motivated us to significantly revise and improve the manuscript. This research has made use of the Extrasolar Planet Encyclopedia, http://exoplanet.eu.
This project has been supported by the Swiss National Science Foundation grant 200020-172505, and has been carried out in part within the frame of the National Centre for Competence in Research (NCCR) PlanetS. 
\end{acknowledgements}

\appendix
\section{Stellar models}
\label{Appendix:stars}

\begin{figure*}[h]
\centering
\includegraphics[width=.5\textwidth, angle=0]{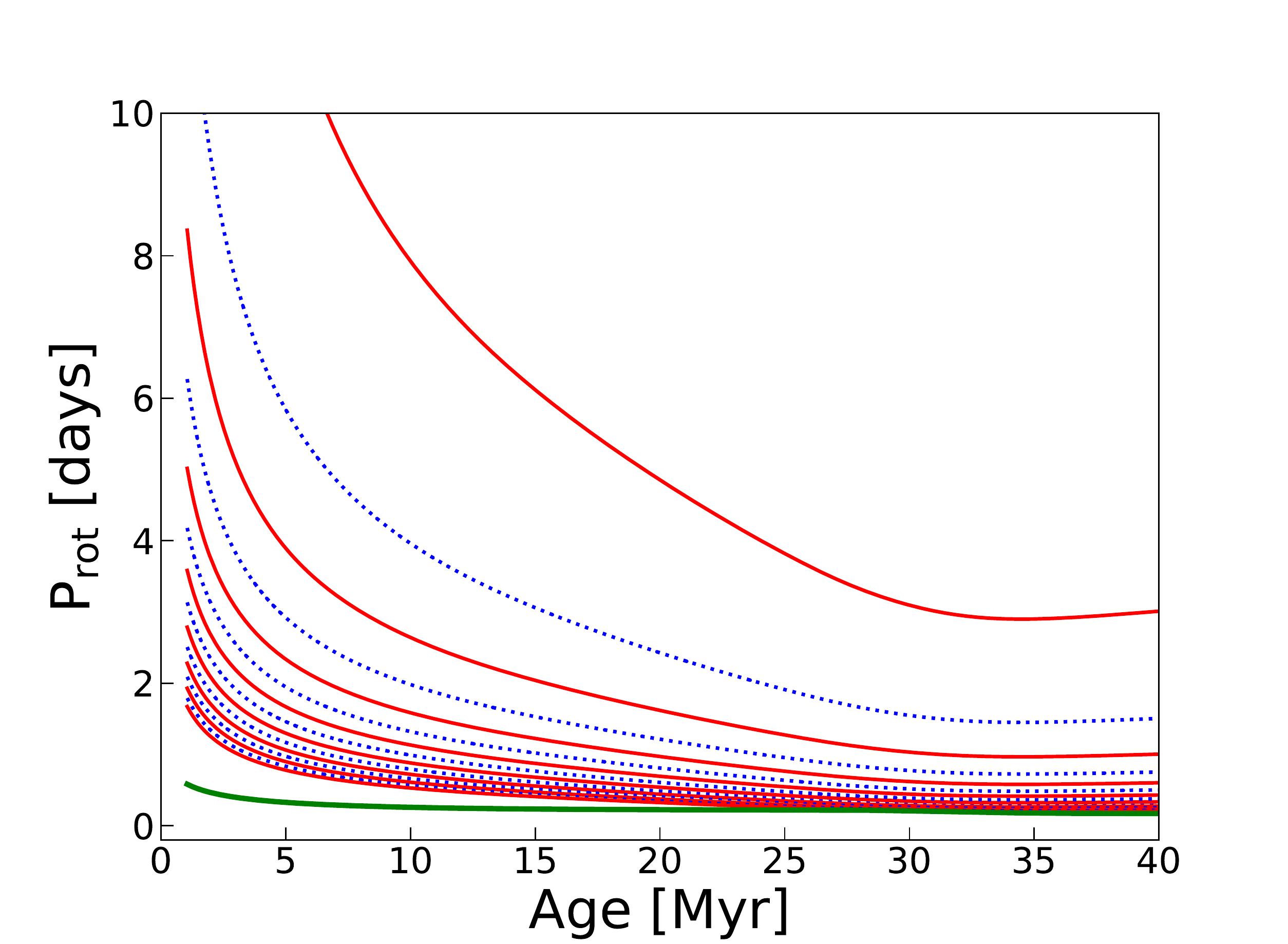}\includegraphics[width=.5\textwidth, angle=0]{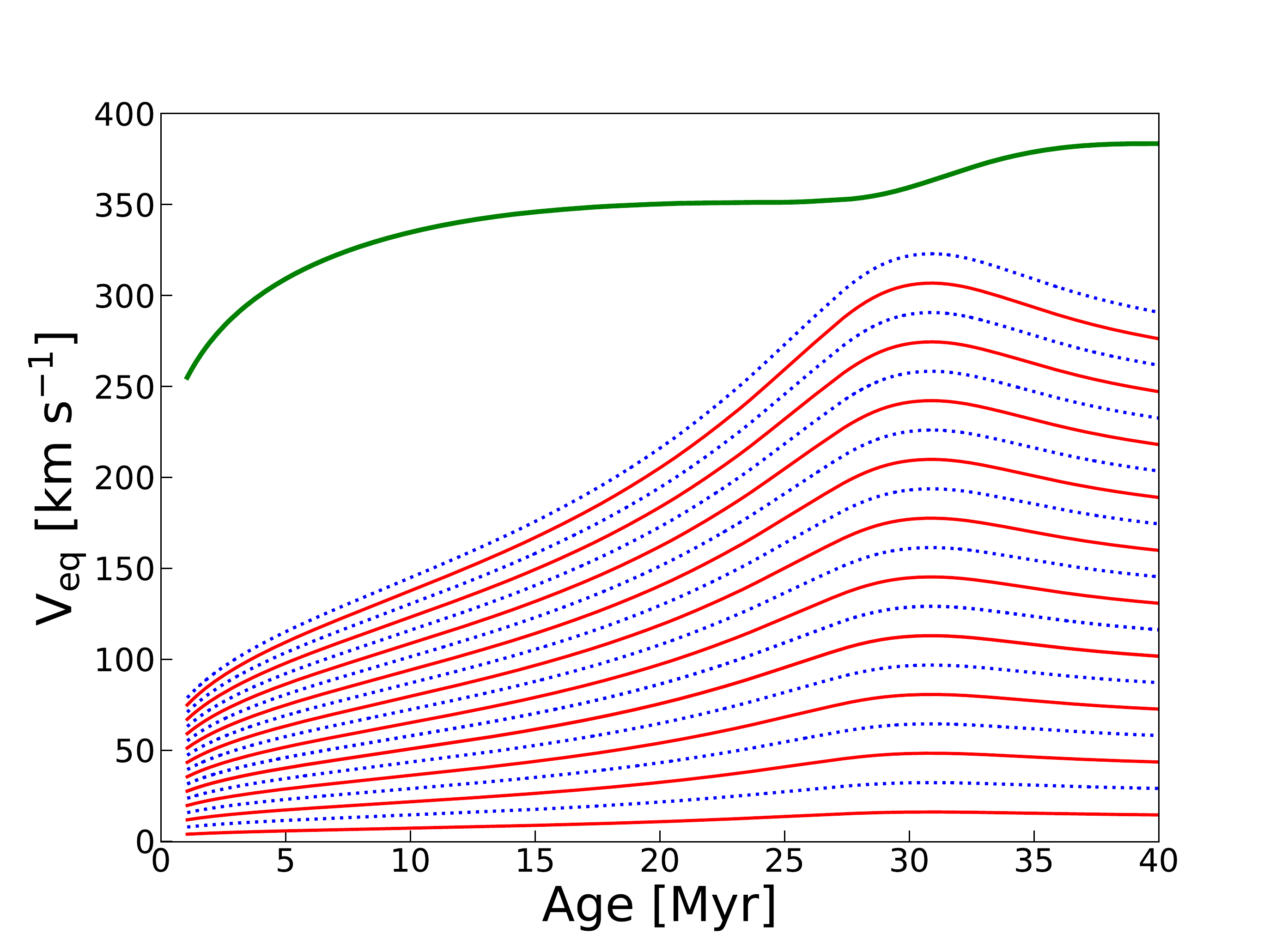}
\caption{Evolution of the rotation period (left panel) and of the surface velocity (right panel) for an isolated 1 M$_\odot$ stellar model during the Pre-Main-Sequence phase and the very early Main-Sequence phase for different values of the initial rotation. The green line shows respectively the critical period (left panel) and the critical velocity (right panel). Note that the bottom line in the left panel corresponds to the upper line in the right panel.}
\label{ROT_1Msol_ISO}
\end{figure*}

Figure~\ref{ROT_1Msol_ISO} shows the evolution of the rotation period and of the surface velocity (at the equator) of a 1 M$_\odot$ stellar model from the Pre-Main-Sequence phase to the early Main-Sequence phase
(only the first 40 Myr are shown).
The star is here assumed to rotate as a solid body. 
During the Pre-Main-Sequence phase, the star is contracting rapidly. This decreases the rotation period (increasing the surface equatorial velocity). The position of the ZAMS, that may be defined here as the time at which the minimum radius is reached, corresponds to the peak in velocity shown on the right panel.
During the Main-Sequence phase, the surface velocity decreases due to the wind magnetic braking. Whatever the rotation of the star on the ZAMS, after 4.6 Gyr, the surface rotation would  be less than 2 km s$^{-1}$, quite
in the range of values observed for the Sun. 

\section{Planet models}
\label{Appendix:orbit}

\begin{figure*}[h]
\centering
\includegraphics[width=.505\textwidth, angle=0]{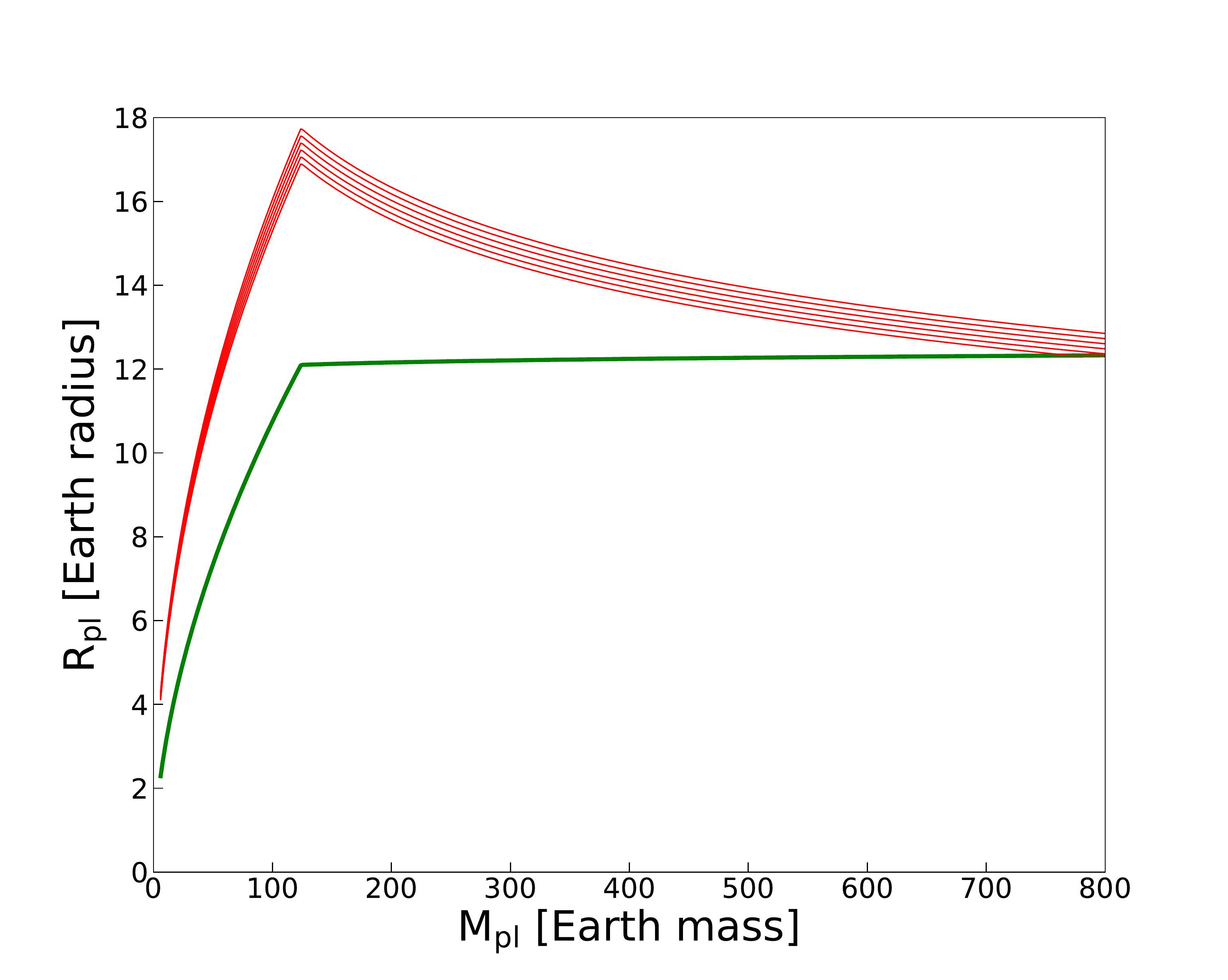}\includegraphics[width=.495\textwidth, angle=0]{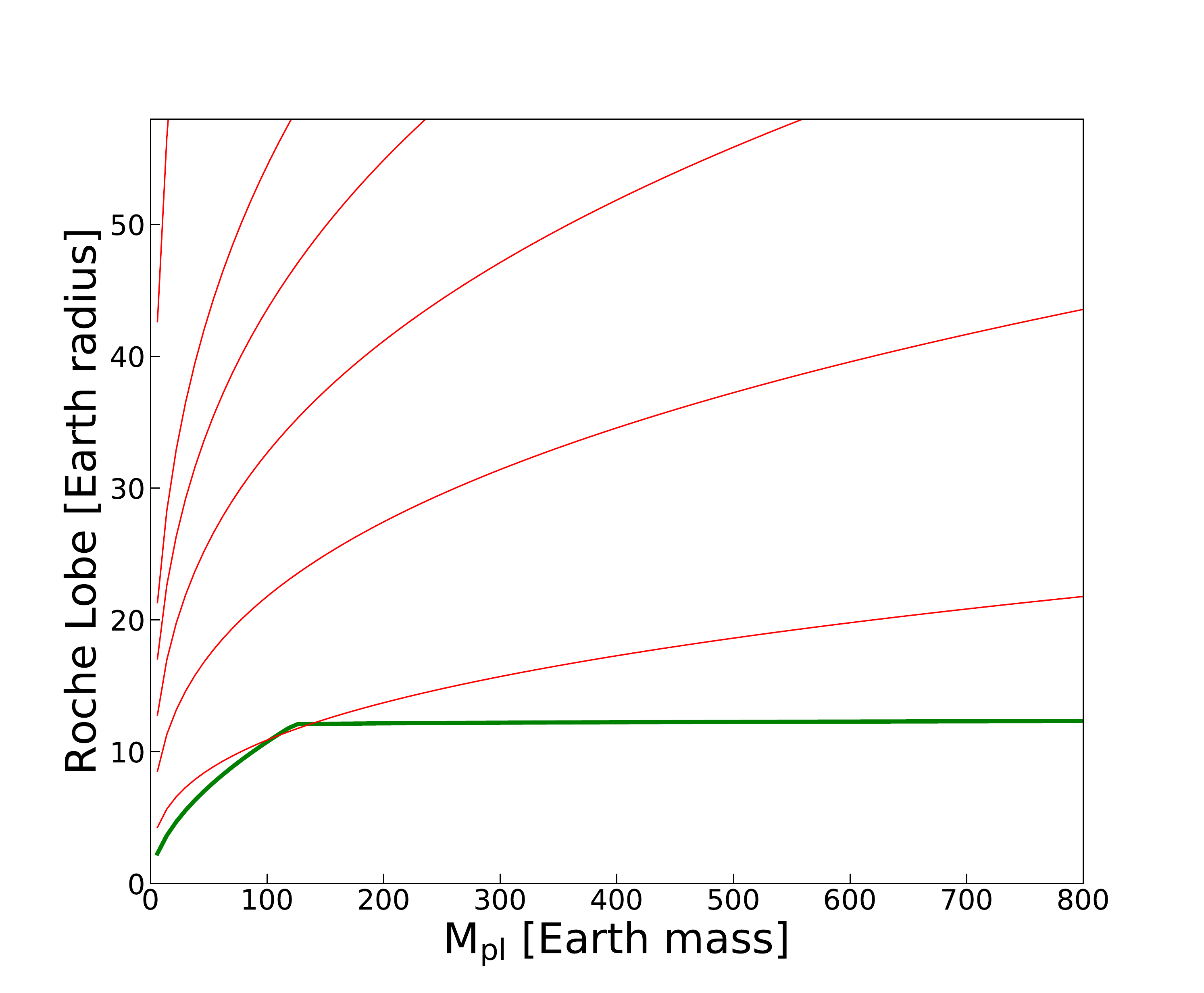}
\caption{{\it Left panel:} Variation of the planet radius as a function of the planet mass (green curve, see text). The red curves indicate the effective radius accounting for the fact the atmosphere of the planet can be inflated. From bottom to top, the curves corresponds to values of the XUV flux between 10$^{4.5}$ and 10$^{5.5}$ in steps of 0.2 dex. These
XUV fluxes corresponds to XUV luminosities equal to 10$^{-3}$ L$_\odot$ and orbital distances between about 0.05 and 0.2 au. 
{\it Right panel:} Variation of the Roche lobe as a function of the planetary mass (red curves) compared to the planetary radius (thick green curve). The red curves corresponds to distances between the planet and the star
equal to 0.01, 0.02, 0.03, 0.04, 0.05 and 0.1 au from bottom to top. The mass of the star is 1 M$_\odot$.
}
\label{Rpl}
\end{figure*}

The variation of the planet radius as a function of the planetary mass, from \citet{bashi2017}, is shown in the left panel of Fig.~\ref{Rpl} (see the green curve). The piece-wise description provides a distinction between {\it small} and {\it large} planets. The transition mass separating the {\it small} planets from the {\it large} planets is ${ m_{ t}}= 124 ~ { m_{ Earth}}$ and the transition radius is ${ r_{ t}}= 12.1 ~ { r_{ Earth}}$. 

When the planet is very close to the star, its atmosphere may be inflated due to heating caused by the XUV irradiation from the star. In that case,
the radius of the planet has to be corrected. The inflated radius is given by $\beta r_{ pl}$, where $r_{ pl}$ is the original planet radius, and $\beta$ taken as in  \citet{Salz2016}. The expression for $\beta$ is valid only 
in a limited domain of gravitational potential, between 10$^{12.2}$
and 10$^{13.6}$ ergs g$^{-1}$. Using the mass-radius relation by \citet{bashi2017}, this corresponds to masses between 6 and 800 Earth masses.

The left panel of Fig.~\ref{Rpl} shows the variation of the original radius (green solid line) and
of the inflated radius (red solid lines) as a function of the mass of the planet for different irradiation values (see caption).

If a blob of planetary matter receives sufficient energy to enable it to reach the limit of the Roche lobe between the planet and the star, then it will escape from the effective gravitational potential well of the planet orbiting its star, and will be lost. The Roche lobe boundary, $R_{ Roche}$, is taken from \citet{erkaev2007}. Its variation as a function of the planet mass for different orbital distances is shown in the right panel of Fig.~\ref{Rpl}.
We see that in general, the Roche lobe is larger than the (non-inflated) planetary radius, except for cases when the orbital distance becomes equal to about 0.01 au or less. In the case XUV absorption radii are considered, orbital distances larger than about 0.02 au are needed to avoid that the Roche lobe becomes smaller than the planet radius.
The present treatment is not sophisticated enough to follow the evolution in a very rigorous way when the planet radius exceeds the Roche lobe, because this effect involves a mass loss mechanism different from the one triggered by the XUV luminosity of the star.
Likely in such cases, the planet will lose its whole atmosphere in a very short timescale, and for gaseous planets, nearly complete evaporation may occur. This is the scenario considered to occur in the present work.

\begin{figure*}[h]
\centering
\includegraphics[width=.5\textwidth, angle=0]{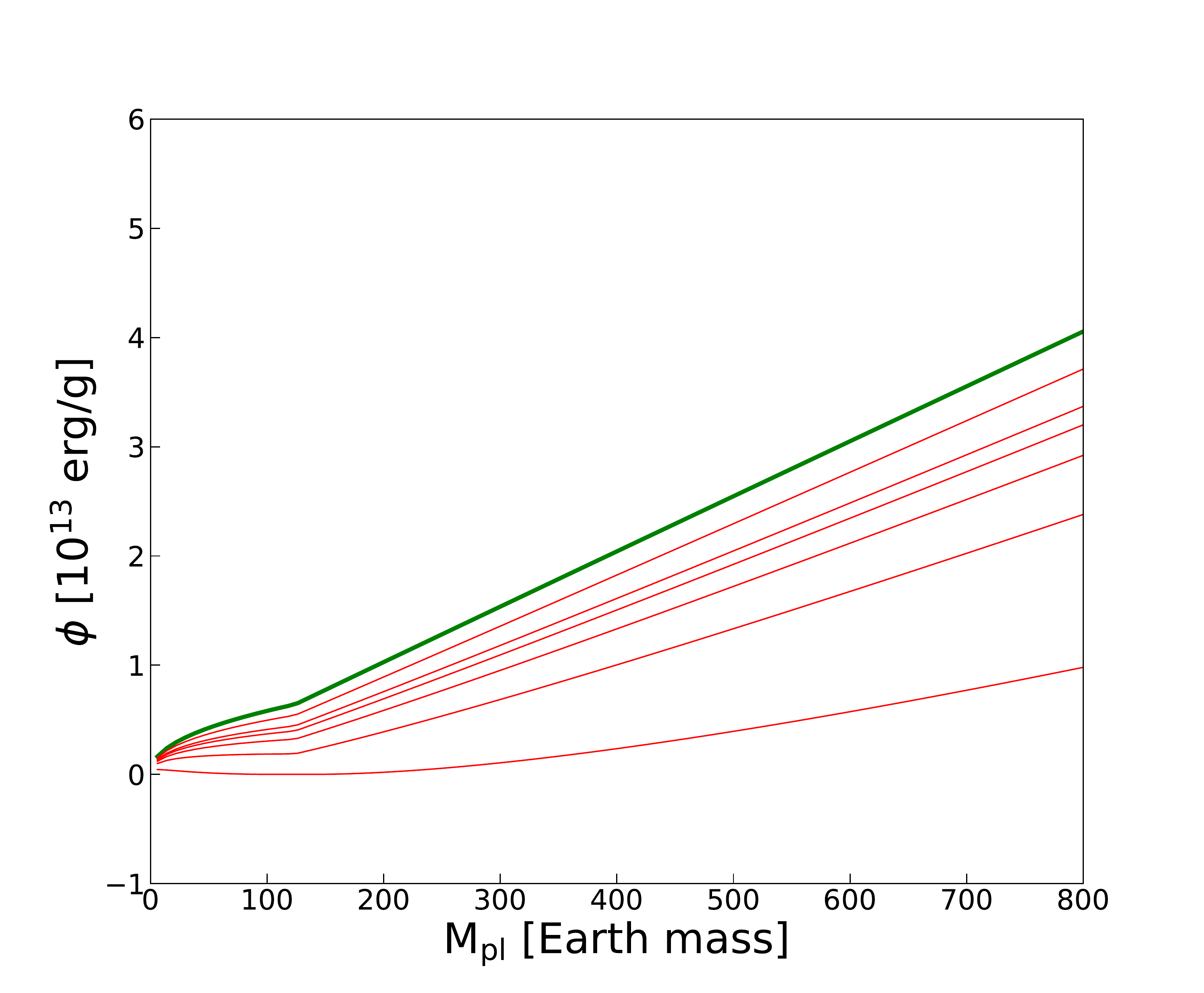}\includegraphics[width=.52\textwidth, angle=0]{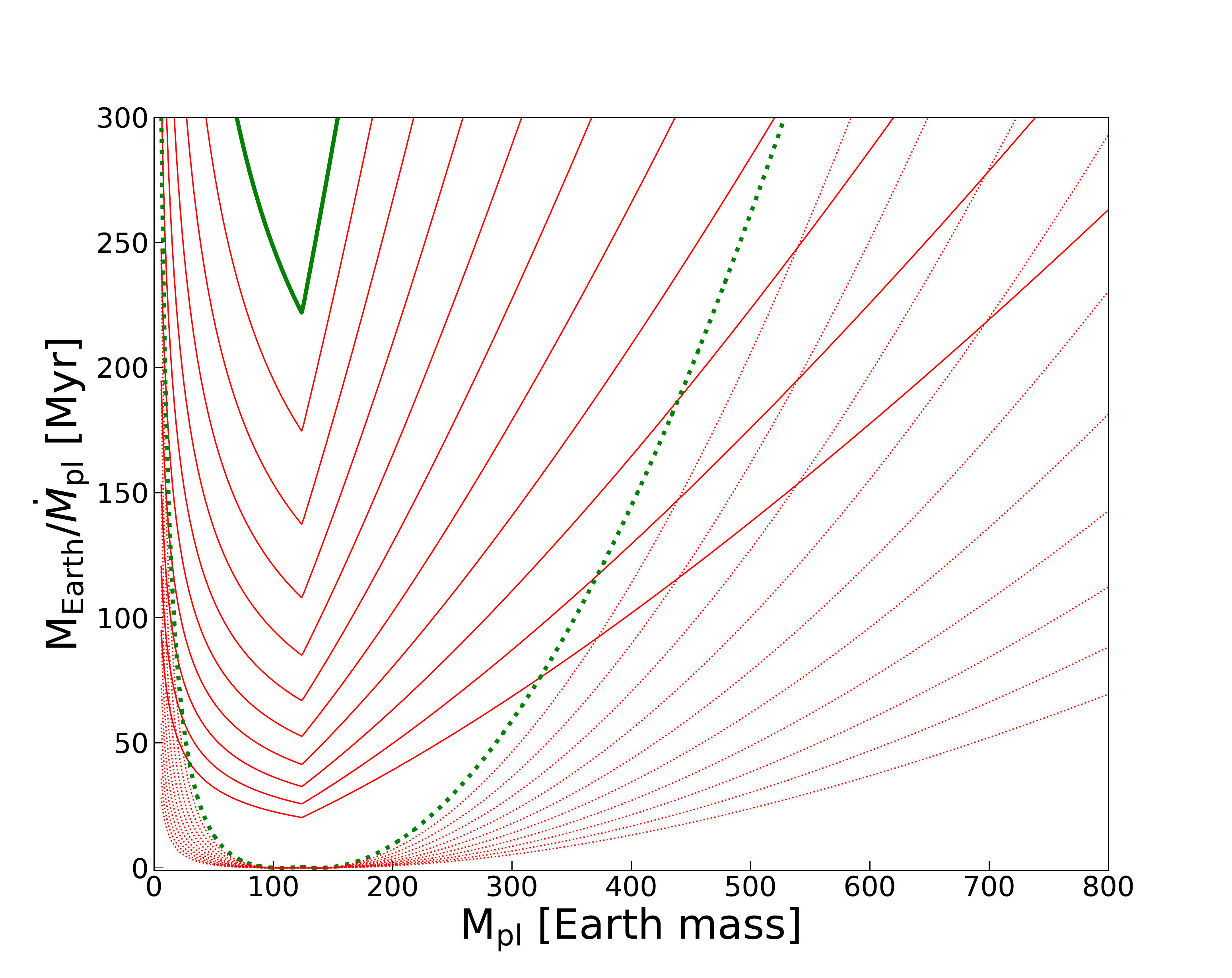}
\caption{{\it Left panel:} Variation of the energy needed to bring a unit mass planet material up to the Roche lobe as a function of the mass of the planet and of the orbital distance. The green line corresponds to the case when the planet is in isolation, showing the energy needed to bring a unit mass to infinity. The red lines corresponds to cases where the planet is sufficiently near the star so that the energy would be lesser than the former case, needed for bringing a unit mass only up to the Roche lobe between the planet and the star. From top to bottom, red lines correspond to orbital distances equal to 0.1, 0.05, 0.04, 0.03, 0.02, and 0.01 au.
{\it Right panel:} The variation of the initial evaporation timescale of planets of different masses irradiated by different XUV fluxes. The green line corresponds to a flux equal to 10$^{4.5}$ ergs s$^{-1}$. The lowest red curve corresponds to a flux equal to 10$^{5.5}$  ergs s$^{-1}$. The lines in between correspond to intermediate values, from 10$^{4.6}$ to 10$^{5.4}$ in steps of 0.1 dex. The distance between the planet and the star has been assumed to be 0.1 au for the continuous lines and 0.01 au for the dashed lines.
}
\label{Pote}
\end{figure*}

In the left panel of Fig.~\ref{Pote}, the energy required for a unit mass to reach the Roche lobe, as given by \citet{erkaev2007}, is shown for different values of the mass of the planet and for different distances to the star (distances decreases from the green line to the lowest red curve).

In the right panel of Fig.~\ref{Pote}, the timescale needed for the evaporation of 1 Earth mass according to the evaporation law given by Eq.~\ref{eq2} is shown as a function of the mass of the planet and for different XUV irradiation fluxes. The most striking
feature is the fact that the evaporation timescale has a minimum where the planet mass-radius relation (shown in the left panel of Fig.~\ref{Rpl}) changes slope.
For masses below the mass corresponding to that transition, the increase of the XUV absorption radius ($\beta r_{ pl}$) overcomes the relatively gentle increase with the planet mass of $K\lvert \phi_{ G}\rvert$. For masses above this mass limit, $\beta r_{ pl}$ decreases, while $\Delta \Phi$ increases, therefore the evaporation timescale increases. Hence, we see that the mass-radius relation plays a key role here. We note that the masses of the planets that
show the largest evaporation rates are masses around that transition (here $\sim$120 M$_{ Earth}, ~ \sim0.38{ M}_{ Jup}$). 

The process of planet evaporation during the Pre-Main-Sequence phase has a rather limited effect when the distance is equal or above 0.1 au. Indeed, the PMS phase lasts 30-40 Myr, while the time for evaporating 1 earth mass from a  100$~M_{ Earth}$ is around 250 Myr. 
When the distance is decreased by a factor 10, the situation is very different. Here we have that whatever the irradiation flux provided superior to 10$^{4.5}$ ergs s$^{-1}$ cm$^{-2}$, the low mass planets with a few $M_{ Earth}$ will be significantly affected by the evaporation process.

\bibliographystyle{aa} 
\bibliography{Evaporation.bib} 

\end{document}